\documentclass[aps, nofootinbib, graphicx]{revtex4-1}
 
\usepackage{hyperref}
\usepackage{amssymb}
\usepackage{amsmath}
\usepackage{verbatim}
\usepackage{mathrsfs}
\usepackage{amsfonts}
\usepackage{latexsym}
\usepackage{epsfig}
\usepackage{epstopdf}
\usepackage[usenames, dvipsnames]{color}
\usepackage{revsymb}
\usepackage{graphicx}
\usepackage{float}
\usepackage{subfig}

\begin{document}


\definecolor{pink}{rgb}{0.9,0,0.6} 

\newcommand{\dario}[1]{\textcolor{red}{{\bf Dario: #1}}}
\newcommand{\claudia}[1]{\textcolor{pink}{{\bf Claudia: #1}}}
\newcommand{\jcd}[1]{\textcolor{blue}{{\bf Juan: #1}}}

\newcommand{\beq}{\begin{equation}}
\newcommand{\eeq}{\end{equation}}
\newcommand{\bea}{\begin{eqnarray}}
\newcommand{\eea}{\end{eqnarray}}
\newcommand{\beqn}{\begin{equation*}}
\newcommand{\eeqn}{\end{equation*}}
\newcommand{\bean}{\begin{eqnarray*}}
\newcommand{\eean}{\end{eqnarray*}}

\renewcommand{\t}{\times}
\long\def\symbolfootnote[#1]#2{\begingroup%
\def\thefootnote{\fnsymbol{footnote}}\footnote[#1]{#2}\endgroup}


\newcommand{\tg}{\tilde{\gamma}}
\newcommand{\tG}{\tilde{\Gamma}}
\newcommand{\tA}{\tilde{A}}
\newcommand{\tR}{\tilde{R}}
\newcommand{\tnabla}{\tilde{\nabla}}

\newcommand{\hg}{\hat{\gamma}}
\newcommand{\hG}{\hat{\Gamma}}
\newcommand{\hA}{\hat{A}}
\newcommand{\hR}{\hat{R}}
\newcommand{\hD}{\hat{\Delta}}
\newcommand{\hnabla}{\hat{\nabla}}

\newcommand{\fg}{\mathring{\gamma}}
\newcommand{\fG}{\mathring{\Gamma}}
\newcommand{\fR}{\mathring{R}}
\newcommand{\fnabla}{\mathring{\nabla}}

\newcommand{\lb}{\pounds_{\vec{\beta}}}

\newcommand{\Lie}[2]{\pounds_{\vec{#1}}{#2}}
\newcommand{\p}{\partial}

\newcommand{\gpar}{{\hat{\gamma}}_\parallel}
\newcommand{\gper}{{\hat{\gamma}}_\perp}

\newcommand{\tend}{t_{\rm end}}
\newcommand{\confH}{\mathcal{H}}

\newcommand{\Mpl}{M_{\rm Pl}}


\title{Gravitational and electromagnetic perturbations of a
charged black hole in a general gauge condition}

\author{Claudia Moreno} 
\email[]{claudia.moreno@cucei.udg.mx} 
\affiliation{Departamento de F\'isica,
Centro Universitario de Ciencias Exactas e Ingenier\'ias, Universidad de Guadalajara\\
Av. Revoluci\'on 1500, Colonia Ol\'impica C.P. 44430, Guadalajara, Jalisco, M\'exico}

\author{Juan Carlos Degollado} 
\email[]{jcdegollado@icf.unam.mx}
\affiliation{Instituto de Ciencias F\'isicas, Universidad Nacional Aut\'onoma 
de M\'exico, Apartado. Postal 48-3, 62210, Cuernavaca, Morelos, M\'exico.}

\author{Dar\'{\i}o N\'u\~nez}
\email[]{nunez@nucleares.unam.mx}
\affiliation{Instituto de Ciencias Nucleares, Universidad Nacional
  Aut\'onoma de M\'exico, Circuito Exterior C.U., A.P. 70-543,
  M\'exico D.F. 04510, M\'exico}

\author{Carlos Rodriguez-Leal} 
\email[]{carlos.rodriguez5050@alumnos.udg.mx} 
\affiliation{Departamento de F\'isica,
Centro Universitario de Ciencias Exactas e Ingenier\'ia, Universidad de Guadalajara\\
Av. Revoluci\'on 1500, Colonia Ol\'impica C.P. 44430, Guadalajara, Jalisco, M\'exico}


\date{\today}


\begin{abstract}
We derive a set of coupled equations for the gravitational and electromagnetic perturbation in the Reissner-Nordstr\"om geometry using the Newman Penrose formalism. 
We show that the information of the physical gravitational signal is contained in the Weyl scalar function $\Psi_4$, as is well known, but for the electromagnetic signal the information is encoded  
in the function $\chi$ which relates the perturbations of the radiative Maxwell scalars $\varphi_2$ and the Weyl scalar $\Psi_3$.
In deriving the perturbation equations
we do not impose any gauge condition and our analysis contains as a limiting case the results obtained previously for instance in Chandrashekhar's book. In our analysis, we also include the sources for the perturbations and focus on a dust-like charged fluid distribution falling radially into the black hole. Finally, by writing the functions on a basis of spin weighted spherical harmonics and the Reissner-Nordstr\"om spacetime in Kerr-Schild type coordinates a hyperbolic system of coupled partial differential equations is presented and numerically solved. In this way, we solve completely a system which generates a gravitational signal as well as an electromagnetic/gravitational one, which sets the basis to find correlations between them and thus facilitating the gravitational wave detection via the electromagnetic signal.
\end{abstract}




\maketitle



\section{Introduction}
\label{sec:introduction}

Gravitational wave astronomy was born in 2015 with the discovery of the first astronomical source named GW150914 \cite{Abbott:2016blz,TheLIGOScientific:2016qqj,Abbott:2016nmj}.
This first observation of a binary black hole system was done 
with the gravitational wave interferometer LIGO \cite{Abbott:2009ij}, which detected 
the gravitational wave signal produced by the
merger of two black holes in a binary system. The two black holes were not surrounded by significant amount of matter which could generate electromagnetic emission and, in this respect, the emission of the merger was purely in the gravitational channel. 
Unlike black hole binaries, a system involving neutron stars do possess electromagnetic counterpart, as
demonstrated by the detection of the neutron star merger event in gravitational and electromagnetic messenger channels in 2017 named GW170817 \cite{TheLIGOScientific:2017qsa, 0067-0049-211-1-7}. 

During the  formation of an accretion disk around a black hole, electrons may escape from the influence of the central object leaving a net charge in the system, these charged particles may be captured by the black hole producing a charged black hole \cite{Zakharov:2014lqa}. 
Thus, the study of the electromagnetic and gravitational perturbations can be used to describe the scattering of both types of waves. However, for a charged black hole a gravitational perturbation of the metric inevitably accompanies a perturbation on the electromagnetic field and vice versa.
The description of coupled electromagnetic and gravitational perturbations on charged black holes have been discussed in several 
works \cite{Moncrief:1974gw,Moncrief:1974ng,Bicak:1980du, Chitre:1976bb, Leaute:1976sn, Chrzanowski:1975wv, Fabbri:1977ux} using different techniques including dispersion of waves due to curvature potentials \cite{Zerilli:1974ai, Ferrari:1984zz, Khanna:2016yow}
and the Newman-Penrose formalism \cite{Chandrasekhar79, Teukolsky:1973ha, Mino:1997bx}.

Chandrasekhar described the 
scattering of electromagnetic waves on a Reissner-Nordstr\"om
black hole and the resulting generation of outgoing gravitational waves, by using the gauge freedom of the Maxwell equations in a curved background,
he derived the electromagnetic equations by finding a gauge that restore the symmetry to the perturbation equations \cite{Chandrasekhar83}. He showed that the curved spacetime produced by a black hole is sensitive to the electromagnetic field part $\chi$ of the spacetime
and this awareness is manifested in the symmetry of the equations for the scalars $\varphi_2^{(1)}$ and $\Psi_3^{(1)}$ in a curved background.
With the 
introduction of this particular gauge the  electromagnetic and gravitational perturbation equations simplify greatly. Furthermore, using this gauge 
the equations for the gravitational and electromagnetic Weyl scalars decouple from the rest of the functions appearing in the system of equations.
Due to the apparent cognizance of the curved geometry to the existence of the Maxwell field in which the symmetry in the equations is recovered, Chandrasekhar dubbed this gauge as \emph{the phantom gauge}.
Since then, 
the scattering of both gravitational and electromagnetic waves have been described using this gauge in a variety of works 
\cite{Lee:1976jp, Lee:1977nc, Chandrasekhar75}.
In Ref~\cite{Lee:1976jp}, Lee found a pair of equations for only two gauge invariant quantities involving electromagnetic and gravitational perturbations in a Kerr-Newman spacetime without using the phantom gauge.
In this work we revisit the findings of Lee including explicitly the matter sources that may cause the perturbation in a Reissner-Nordstr\"om background.

As a direct application of our setting we
consider a pressure-less charged perfect fluid (dust) as the cause of the perturbation falling radially into the black hole.
We write explicitly the equations in a coordinate system
and expand the functions using a spherical harmonic basis with the appropriate spin weight. With this choice, we show that the dynamics of the perturbed functions $\Psi_4^{(1)}$ and $\chi$, is completely determined by a set of partial differential equations that depend only on the radial and temporal coordinates. We numerically solve such set of equations and obtain the wave forms for several representative values of the parameters of the system. Setting the basis for a thorough comparatively analysis which might find correlations between the waveforms and thus, by a detection of one of these electromagnetic/gravitational signal, one will be able to infer the presence of a purely gravitational one. This program will be carried out in future works.

The paper is structured as follows:
In section \ref{sec:foundations} we introduce the basics of the Newman Penrose formalism including the Bianchi and Maxwell identities.
In section \ref{sec:Perturbations} we provide a detailed derivation of the the perturbation equations in a Reisnner-Nordstr\"om bakground.
In section \ref{sec:sources} we describe the sources of the perturbation and show that choosing an adequate decomposition in spin weighted spherical harmonics it is possible to separate the time-spatial structure of the equations. 
In Section \ref{sec:PME} we introduce the tetrad and geometric quantities in the Reissner–Nördstrom background described in horizon penetrating coordinates, present a numerical scheme to solve the equations for some particular scenario of matter falling into the black hole, and show some waveforms, gravitational as well as those related with $\chi$. Finally some concluding remarks are given in section \ref{sec:conclusions}.
In the rest of the paper we use $\eta$ to indicate the signature of the metric: $\eta=1$ for signature $(+,-,-,-)$, and $\eta=-1$ for the signature $(-,+,+,+)$, and we will use geometric units where $c=G=1$.


\section{Foundations: Newman-Penrose formalism}
\label{sec:foundations}

The starting point in the Newman Penrose formalism is to define a tetrad of null vectors~\cite{Newman:1962a}.
The choice of the tetrad is made to reflect symmetries of spacetime, since certain components may
vanish, leading to simplification of field equations.
In this work we use  $l^\mu$ and $k^\mu$, ingoing and outgoing null vectors respectively, which satisfy the normalization conditions
$
k_{\mu}l^{\mu}=-1$ and
$
m_{\mu}\overline{m}^{\mu}=1.
$
The metric tensor can be represented by $g_{\mu \nu}=-2(l_{(\mu} k _{\nu_)}-m_{(\mu} \overline m _{\nu)} )$ where $\overline m$ means complex-conjugate, Greek index runs from $0$ to $3$. 
The directional derivative operators are defined as ${\bf D}=l^\mu\,\partial_\mu, {\bf \Delta}=\kappa^\mu\,\partial_\mu$ and ${\bf \delta}=m^\mu\,\partial_\mu$. The spin coefficients are obtained from the projections 
\begin{eqnarray}
\kappa&=&m^\mu\,l_{\mu;\nu}\,l^\nu; \hspace{0.6cm} \tau=m^\mu\,l_{\mu;\nu}\,k^\nu; 
\hspace{0.5cm} \sigma=m^\mu\,l_{\mu;\nu}\,m^\nu; \hspace{0.7cm} 
\rho=m^\mu\,l_{\mu;\nu}\,{\overline m}^\nu; \nonumber \\
\hat \pi&=&-\overline m^\mu\,{k}_{\mu;\nu}\,l^\nu; \hspace{0.5cm} 
\nu=-\overline m^\mu\,{k}_{\mu;\nu}\,k^\nu; 
\hspace{0.5cm} \mu=-\overline m^\mu\,{k}_{\mu;\nu}\,m^\nu; 
\hspace{0.5cm} \lambda=-\overline m^\mu\,{k}_{\mu;\nu}\,{\overline m}^\nu; \nonumber \\
\epsilon&=&\frac12\left(k^\mu\,l_{\mu;\nu} - m^\mu\,{\overline 
m}_{\mu;\nu}\right)\,l^\nu; 
\hspace{0.5cm} \gamma=\frac12\left(k^\mu\,l_{\mu;\nu} - \overline m^\mu\,{ 
m}_{\mu;\nu}\right)\,k^\nu;\nonumber \\
\beta&=&\frac12\left(k^\mu\,l_{\mu;\nu} - \overline m^\mu\,{ 
m}_{\mu;\nu}\right)\,m^\nu; 
\hspace{0.5cm} \alpha=\frac12\left(k^\mu\,l_{\mu;\nu} - 
\overline m^\mu\,{m}_{\mu;\nu}\right)\,{\overline m}^\nu, 
\label{eq:coef_spinors}
\end{eqnarray} 
where ";" stands for covariant derivative. The Weyl scalars related with the curvature $\Psi_0$ $\Psi_1$, $\Psi_2$, $\Psi_3$ and $\Psi_4$, and the source terms $\Phi_{ij}$,
are defined as
\begin{eqnarray}
\Psi_0&=&-C_{\mu\nu\lambda\tau}\,l^\mu\,{m}^\nu\,{l}^\lambda\,{m}^\tau=-C_{l m l m}; \hspace{1.3cm} \Psi_1=-C_{\mu\nu\lambda\tau}\,l^\mu\,k^\nu\,{l}^\lambda\,{m}^\tau=-C_{l k l m}; \nonumber \\\Psi_2&=&-C_{\mu\nu\lambda\tau}\,l^\mu\,{m}^\nu\,\overline{m}^\lambda\,{k}^\tau=-C_{l m \overline m k}; \hspace{1.15cm} \Psi_3=-C_{\mu\nu\lambda\tau}\,l^\mu\,k^\nu\,\overline{m}^\lambda\,{k}^\tau=-C_{l k \overline m k}; \nonumber \\
\Psi_4&=&-C_{\mu\nu\lambda\tau}\,k^\mu\,\overline{m}^\nu\,k^\lambda\,\overline{m}^\tau=-C_{k \overline m k \overline m}, \label{eqs:Weyl_sc} \end{eqnarray}
and
\begin{align}
\Phi_{00} =\overline \Phi_{00} = 4 \pi T_{\mu \nu} l^\mu l^\nu \equiv 4\pi T_{ll}; & 
\qquad \Phi_{01} = {\overline \Phi_{10}}= 4 \pi T_{\mu \nu} l^\mu m^\nu \equiv 4\pi T_{lm};   \nonumber \\
\Phi_{02} = {\overline \Phi_{20}} = 4 
\pi T_{\mu \nu} m^\mu m^\nu \equiv 4\pi T_{mm}; &  \qquad  \Phi_{11} =2\pi T_{\mu \nu} ( l^\mu k^\nu+  m^\mu \overline m^\nu)\equiv 2\pi \left(T_{lk}+T_{m \overline{m}}\right); \nonumber \\
\Phi_{22} =\overline \Phi_{22} = 4 \pi 
T_{\mu \nu} k^\mu k^\nu \equiv 4\pi T_{kk}; & \qquad
\Phi_{12} =\overline \Phi_{21} = 4 \pi 
T_{\mu \nu} k^\mu m^\nu \equiv 4\pi T_{km},
\label{proy_faraday}
\end{align}
where $C_{\mu\nu\lambda\tau}$ is the Weyl tensor and $T_{\mu\nu}$ is the stress energy tensor of the matter content. \\
The information of the electromagnetic fields is encoded in the scalars,
\begin{equation}
\varphi_0=F_{\mu\nu}l^{\mu}m^{\nu}; \hspace{0.5cm} \varphi_1=\frac{1}{2} F_{\mu\nu}(l^{\mu}k^{\nu} +
\overline m^{\mu} m^{\nu}); \hspace{0.5cm} \varphi_2=F_{\mu\nu}\overline m^{\mu}k^{\nu},
\label{eq:varphis}
\end{equation}
where $F_{\mu\nu}$ is the Faraday tensor \cite{Chandrasekhar75}.
The stress energy tensor for the electromagnetic field has the form
\begin{equation} \label{eq:tmunu}
T_{\mu\nu} = \frac{\eta}{4\pi\mu_0}\,\left( F_{\mu\lambda}\,{F^\lambda}_{\nu} + \frac{1}{4}g_{\mu\,\nu}F^{\alpha\beta}\,F_{\alpha\beta}
\right),
\end{equation}
with $\mu_0$ the magnetic permeability in vacuum. In the rest of this work we set $\mu_0=1$. \\ 
From the definition of the scalars $\Phi_{ij}$ and Eqs.\eqref{eq:varphis} and \eqref{eq:tmunu} we consider that $
\Phi_{ij} =2\,\eta\, \varphi_{i}\,\overline\varphi_{j}. $
Latin index runs from $1$ to $3$. In order to obtain the electromagnetic and gravitational perturbation equations, we depart from the  projected Maxwell equations, and the Bianchi identities \cite{Chandrasekhar83,Degollado:2011gi} as explained below.

\subsection{Maxwell equations}

The dynamics of the electromagnetic fields with sources is described by the Maxwell equations $F^{\mu \nu}{}_{;\nu}=\,J^\nu$, where $J^{\nu}$ is the external electric current.
The Maxwell equations projected along the tetrad $\eta^{pn}\,F_{ap\,|n}=J_a$ ("$_|$" means intrinsic derivative),
written in terms of the spin coefficients and the electromagnetic scalars are \cite{Degollado:2009rw} 
\begin{eqnarray}
(D-2 \eta \rho) \varphi_1 - (\overline{\delta} + \eta(\pi - 2 \alpha)) 
\varphi_{0}+\eta \kappa \varphi_2= \frac{\eta}{2}\,J_l, \label{Muno} \\
(\delta-2 \eta \tau) \varphi_1 - (\Delta+ \eta(\mu- 2 \gamma)) 
\varphi_{0}+\eta \sigma \varphi_2= \frac{\eta}{2}\,J_m, \label{Mdos} \\
(D- \eta(\rho - 2 \epsilon)) \varphi_2 - (\overline{\delta} + 2 \eta \pi ) \varphi_{1}+\eta \lambda \varphi_0= 
\frac{\eta}{2}\,J_{\overline{m}}, \label{Mtres} \\
(\delta-\eta(\tau + 2 \beta)) \varphi_2 - ( \Delta + 2 \eta \mu) \varphi_{1}
+\eta \nu \varphi_0=\frac{\eta}{2}\,J_k, \label{Mcuatro}
\end{eqnarray}
where
$J_l = J_{\mu} l^{\mu}, J_m = J_{\mu} m^{\mu}$ and $J_k = J_{\mu} k^{\mu}$.  

\subsection{Bianchi identities}

By projecting the equations
$R_{\mu\nu\,\left(\lambda\tau;\sigma\right)}=0$ on the tetrad, one gets a set of equations which include the spin coefficients and operators acting on the Weyl scalars
known as the Bianchi identities.
In the following we will use two of them, Eq. (321d) and Eq. (321h) from \cite{Chandrasekhar83}: 
\begin{eqnarray}
&&(D+ \eta\,(4 \epsilon\, -\rho\,) )\Psi 
_{4}-(\overline{\delta}+2\eta\,(2\pi\, +\alpha\,) )\Psi _{3}
+\left(3\eta\, \Psi _{2}+2\Phi _{11}\right)\,\lambda \nonumber \\
&&=\eta\,(\overline{\delta}+2\eta\,(\alpha\, - \overline \tau))\Phi _{21}
-\eta\, \left( \Delta +\eta\,(\overline{\mu} + 2 \gamma\, - 2 \overline 
\gamma\, )\right)\Phi _{20}  +2\,\nu \, \Phi_{10}+\overline \sigma \, \Phi_{22}  \label{R321dB0_0}, \\ 
&& \nonumber \\
&&-(\delta +\eta\,(4\beta - \tau) )\Psi _{4}+(\Delta 
+ 2\eta\,( \gamma\, +2\mu\,) )\Psi _{3}-\left( 3 \eta\, \Psi_{2}-2\Phi 
_{11}\right) \nu  \nonumber \\
&&=\eta\, (\Delta+2\eta\,(\overline{\mu}+\gamma\,))\Phi _{21}
-\eta\,(\overline{\delta}+\eta\,(-\overline \tau+2\alpha\, 
+2\overline{\beta}))\Phi _{22}- \overline \nu \, \Phi_{20} + 2 \, \lambda \, \Phi_{12}. \label{R321hB0_0} 
\end{eqnarray}

For a detailed description on the projections we refer the reader to \cite{Degollado:2011gi}. 
Similarly, projecting the Weyl tensor on the tetrad, we obtain the expression:
%
\begin{equation}
\Psi_4 + \left(\Delta + \eta\,\left(\mu+\overline{\mu} + 3\,\gamma - \overline{\gamma}\right)\right)\,\lambda - \left(\overline{\delta} + \eta\,\left(3\,\alpha + \overline{\beta} + \pi - \overline{\tau}\right)\right)\,\nu=0,
\label{eq:Psi4_0}
\end{equation}
which are the three equations needed in the forthcoming derivation.

\section{Equations for the perturbations}
\label{sec:Perturbations}

Corresponding to the six parameters of the Lorentz group of transformation, there are
six degrees of freedom to rotate a chosen tetrad frame. It is usual to encode a general Lorentz transformation in terms of the basis vectors $l$, $k$ and $m$, and classify them in three classes of rotations, each one of them leaving invariant a vector under the transformation.
The effect of the basis transformation on the various Newman Penrose quantities can be found in \cite{Chandrasekhar83}. 
\begin{eqnarray}
\Psi_0 \rightarrow \Psi_0 ,
\qquad 
\Psi_1 \rightarrow \Psi_1 + a \Psi_0^{(1)} \ ,
\qquad 
\Psi_2 \rightarrow \Psi_2+2 a\Psi_1^{(1)} \ ,
\nonumber \\
\Psi_3 \rightarrow \Psi_3 + 3a\Psi_2^{(1)} \ ,
\qquad 
\Psi_4\rightarrow \Psi_4 + 4a \Psi_3^{(1)} \ ,
\end{eqnarray}
and
\begin{eqnarray}
\varphi_0 \rightarrow \varphi_0  \ ,
\qquad
\varphi_1 \rightarrow \varphi_1 + a\varphi_0^{(1)} \ ,
\qquad
\varphi_2 \rightarrow \varphi_2+2 a\varphi_1^{(1)},
\end{eqnarray}
where the superscript denotes a perturbed quantity.
For the Reissner-Nordstr\"om spacetime the background scalars $\Psi_2$ and $\varphi_1$ are the only non zero. Consequently $\Psi_0$, $\Psi_1$, $\Psi_2$, $\Psi_4$, $\varphi_0$ and $\varphi_1$ are unaffected to first order under an infinitesimal rotation. 
However, $\Psi_3$ and $\varphi_2$ are indeed affected since $\Psi_2$ and $\varphi_1$ are different from zero in the background. Nevertheless, the radiative combination 
\begin{equation}
\chi =\eta\,\left(2\,\varphi_1\,\Psi_3^{(1)} - 3\,\Psi_2\,\varphi_2^{(1)}\right), \label{eq:chi}
\end{equation}
is invariant to first order and, as it is shown below, one can get a coupled system of equations for this function, $\chi$ and the perturbed Weyl component $\Psi_4^{(1)}$.
In the following we describe in detail the general procedure to find such coupled set of equations.

\subsection{Perturbed Maxwell Equations}
\label{sec:PME}
The ingoing and outgoing electromagnetic radiation is given by the perturbations of the scalars $\varphi_0$ and $\varphi_2$ respectively. However, in a charged spacetime the outgoing electromagnetic perturbations couples with the perturbations of the Weyl scalar $\Psi_3$ that carries the so called electromagnetic part of the spacetime. A similar coupling occurs with the ingoing perturbation $\varphi_0$, and $\Psi_1$. In this section we will derive an equation for 
the perturbation of $\varphi_2$.

First, consider the following identities relating the derivative operators \cite{Chandrasekhar83}:
%
\begin{eqnarray}
[\Delta, \overline{\delta} ] &=& \eta \nu D + \eta (\alpha + \overline{\beta} - 
 \overline{\tau}) \Delta - \eta (\overline{\mu}-\overline{\gamma}+ \gamma)\overline \delta - \eta \lambda \delta, 
 \label{eq:con_Dd} \\
\eta \Delta \pi &=& \eta D\nu -\mu(\pi+\overline{\tau}) - \lambda(\overline{\pi}+\tau) - 
\pi (\gamma-\overline{\gamma})+\nu (3\epsilon+\overline{\epsilon})- \eta \Psi_3 - \Phi_{21},  \\
\eta \overline{\delta} \mu&=&\eta \delta \lambda -\nu(\rho-\overline{\rho})-\pi(\mu-\overline{\mu})-
\mu(\alpha+\overline{\beta})-\lambda(\overline{\alpha}-3\beta) + \eta \Psi_3-\Phi_{21}.
\end{eqnarray}
Second, on the projected Maxwell's Eqs. (\ref{Muno}-\ref{Mcuatro}) operate with $[\Delta +\eta(\overline{\mu}-\overline{\gamma}+ \gamma + 2\mu)]$ on 
Eq. (\ref{Mtres}) and with  $[\overline \delta-\eta(\overline{\tau} - \alpha - \overline{\beta} - 2 \pi)]$ on Eq. (\ref{Mcuatro}), then subtract these equations and after some algebra one arrives to the following expression: 
\begin{eqnarray}
&&[ (\Delta+ \eta(\gamma-\overline{\gamma} +2\mu +\overline{\mu}))(D-\eta(\rho_s+2\epsilon))-
(\overline{\delta} + \eta(\alpha +\overline{\beta} + 2 \pi_s - \overline{\tau}))(\delta -\eta(\tau + 2\beta))] \varphi_2 + \nonumber \\
&&- [\eta \nu\,D\,\varphi_1-\eta \lambda \delta\,\varphi_1 + 2\,\varphi_1\,\left((D +\eta(3 \epsilon+\overline{\epsilon} + \rho
- \overline{\rho}_s))\eta \nu
-( \delta +\eta(\overline {\pi} + \tau -\overline{\alpha}+3\beta))\eta \lambda-4\eta \Psi_3\right)]  + \nonumber  \\
&&\varphi_0\,[(\Delta +\eta(\overline{\mu}-\overline{\gamma}+ \overline{\gamma} + 2\mu))\eta \lambda-(\overline{\delta}-\eta(\overline{\tau}
-\alpha 
- \overline{\beta}-2\pi))\eta \nu] +\eta \lambda\,
\Delta\,\varphi_0 -\eta \nu\,\overline{\delta}\,\varphi_0=\frac{\eta}{2}\,J_2, \label{eq:phi2a}
\end{eqnarray}
where
\begin{eqnarray}
J_2&=&(\Delta+\eta(\gamma-\overline{\gamma} +2 \mu + \overline{\mu}))J_{\overline {m}}
-(\overline{\delta} +\eta(\alpha +\overline{\beta} + 2 \pi 
-\overline{\tau})) J_k. 
\end{eqnarray}
In order to describe the perturbation of $\varphi_2$ let us perform a first order perturbation of the form $f\rightarrow f+ 
f^{(1)}$  in all the functions on Eq. \eqref{eq:phi2a}. 
In a Reissner-Nordstr\"om like background, considering that the spacetimes are type $D$, in Petrov classification, the spinor
quantities $\nu, \lambda, \kappa,\sigma$, are zero. Furthermore, we will consider spherical symmetry and that $\varphi_0, \varphi_2$ are zero as well. These considerations imply 
\begin{eqnarray}
&&[ (\Delta+\eta(\gamma-\overline{\gamma} +2\mu +\overline \mu))(D-\eta(\rho+2\epsilon))-
(\overline{\delta} + \eta(\alpha +\overline{\beta} + 2 \pi - \overline{\tau}))(\delta -\eta(\tau + 2\beta))]\varphi_2^{(1)} 
+ \nonumber \\
&&- \left\{\eta {\nu}^{(1)}\,D\,\varphi_1-\eta {\lambda}^{(1)} \delta\,\varphi_1 + 2\,\varphi_1\,\left[(D +\eta (3 \epsilon
+\overline{\epsilon} + \rho - \overline{\rho}))\eta {\nu}^{(1)}
- ( \delta +\eta (\overline{\pi} + \tau -\overline{\alpha}+3\beta))\eta{\lambda}^{(1)} - \nonumber \right. \right. \\
&& \left. \left. 4\eta \Psi_3^{(1)}\right] \right\} 
=\frac{\eta}{2}\,J_2^{(1)}, \label{eqelectro}
\end{eqnarray}
with the perturbed current term defined as
\begin{eqnarray}
J_2^{(1)}&=&(\Delta+\eta(\gamma-\overline{\gamma} +2 \mu + \overline{\mu}))J_{\overline{m}}^{(1)}
- (\overline{\delta} +\eta(\alpha +\overline{\beta} + 
2 \pi -\overline{\tau})) J_k^{(1)}. 
\end{eqnarray}
In the previous equation we have kept only first order and non-vanishing background quantities. Eq. \eqref{eqelectro} can be further simplified by using background Maxwell equations Eqs. 
(\ref{Muno}); $D\varphi_1=2\,\eta \rho\,\varphi_1$, 
and Eq. (\ref{Mdos}); $\delta \varphi_1=2\,\eta \tau\,\varphi_1$
to obtain
\begin{eqnarray}
&&[(\Delta + \eta(\overline{\mu} - \overline{\gamma}+\gamma + 2\mu))(D-\eta(\rho-2\epsilon)) - 
(\overline{\delta} - \eta(\overline{\tau} -\alpha - \overline{\beta}- 2 \pi))(\delta -\eta(\tau -2\beta))]\varphi_2^{(1)} = \nonumber \\
&&2 \eta \varphi_1[(D +\eta( 3 \epsilon+\overline{\epsilon} + 2\rho - \overline{\rho})){\nu}^{(1)}
- (\delta+ \eta( \overline{\pi}  -\overline{\alpha}+3\beta+2\tau)){\lambda}^{(1)} - 
2 \Psi^{(1)}_3] + \frac{\eta}{2}\,J_2^{(1)}.
\label{eq:maxphi2}
\end{eqnarray}
This equation can be written in a more convenient form
using the commutation expressions for the operators 
acting on $\varphi_2^{(1)}$ 
\begin{eqnarray}
\left[ \Delta, D \right] &=& \eta(\gamma +\overline{\gamma})D+\eta(\epsilon + \overline{\epsilon}) \Delta - 
\eta(\overline{\tau} + \pi)\delta -\eta(\tau +\overline{\pi})\overline{\delta}, 
 \\ 
\left[ \overline{\delta}, \delta \right] &=& \eta(\overline{\mu} -\mu)\,D+\eta(\overline{\rho}-\rho)\Delta+
\eta(\alpha-\overline{\beta})\delta+\eta({\beta}-\overline{\alpha})\overline{\delta}, 
\end{eqnarray}
 and the Ricci identities \cite{Moreno:2016urq}:
\begin{align}
\eta D \mu&= \eta \delta \pi +(\overline{\rho} \mu +\sigma \lambda)+\pi (\overline{\pi}-\overline{\alpha}+\beta)-
\mu (\epsilon+\overline{\epsilon})-\nu \kappa +\eta \Psi_2 +2R, \\
\eta \overline{\delta}\,\beta&= -\eta \delta \alpha-(\mu \rho-\lambda \sigma)-\alpha \overline{\alpha} -
\beta (\overline{\beta} -2\alpha) 
-\gamma(\rho-\overline{\rho})-\epsilon(\mu -\overline{\mu}_s)
+\eta \Psi_2-\Phi_{11}-R,
\end{align}
%
on Eq. \eqref{eq:maxphi2}.
After some algebra
one arrives to the following expression relating the perturbations
$\varphi_2^{(1)}, \, {\nu}^{(1)}, \, {\lambda}^{(1)}, \, \Psi_3^{(1)},$ with the current $J_2^{(1)}$ 
\begin{small}
\begin{eqnarray}
&&[(D-\eta\,\left(\rho-3\epsilon-\overline{\epsilon} \right))(\Delta + \eta\,\left(2\mu + \overline{\mu} + 2 \gamma \right)) - 
(\delta + \eta\,\left(3\beta-\overline{\alpha} - \tau\right))(\overline{\delta} + \eta\,\left(2 \alpha
+ 2\pi - \tau\right))-3 \eta \Psi_2]\varphi_2^{(1)} 
= \nonumber \\
&&2 \eta \varphi_1[(D + \eta\,\left(2\rho - \overline{\rho} + 3 \epsilon+\overline{\epsilon} \right)){\nu}^{(1)}
- (\delta - \eta\,\left(\overline{\alpha}-3\beta-\overline{\pi}-2\tau_s\right)){\lambda}^{(1)} - 2 
\Psi^{(1)}_3] + \frac{\eta}{2}\,J_2^{(1)}.
\label{eq:maxphi2c}
\end{eqnarray}
\end{small}
In order to simplify the notation let us define the derivative operators:
\begin{eqnarray}
{\bf D}_N= D + \eta\,N\,\rho + 4\,\eta\,\epsilon, &\hspace{1cm}& {\bf \Delta}_N=\Delta + \eta\,N\,\mu, \label{eq:OpsD_Del}\\
{\bf \delta}_{N}= \delta + \eta\,N\,\beta,  &\hspace{1cm}& {\bf \overline{\delta}}_{N}= \overline{\delta} + \eta\,N\,\beta, \label{def:deltas}
\end{eqnarray}
{\bf where $N$ takes integer values}. In terms of these expressions we can rewrite the electromagnetic perturbation equation Eq. (\ref{eq:maxphi2c}) as:
\begin{equation}
{\bf D}_{+1}\,{\nu}^{(1)} - {\bf \delta}_{4}\,{\lambda}^{(1)} =
\frac{1}{2\,\eta\,\varphi_1}\,\left[{\bf D}_{-1}\,{\bf \Delta}_{3} - {\bf \delta}_{4}\,{\bf \overline{\delta}}_{-2} - 
6\,\eta\,\Psi_2\right]\,{\varphi_2}^{(1)} + 2\,\Psi_{3}^{(1)} + \frac{1}{4\,\varphi_1}\,J_2^{(1)}\ . \label{eq:maxphi3c}    
\end{equation}
With the notation of Eqs.~(\ref{eq:OpsD_Del}, \ref{def:deltas}) we have denoted ${\bf D}_{+1}=D+\eta\,\left((+1)\,\rho + 4\,\epsilon\right)$ for $ N=+1, \, {\bf D}_{-1}=D+\eta\,\left((-1)\,\rho + 4\,\epsilon\right)$ for $N=-1$, and analogously for other values and operators.

\subsection{Perturbed Bianchi identities}

In order to derive the equations for the perturbations $\Psi_4^{(1)}$ and $\Psi_3^{(1)}$ 
we start by perturbing the the Bianchi identities Eqs. (\ref{R321dB0_0}, \ref{R321hB0_0}). As mentioned above, in a Petrov
 type D space-time background, 
 the background Weyl scalars  vanish except $\Psi_2$, and the spinor coefficients $\kappa,\nu, \lambda, \sigma$, are equal to zero. Furthermore, we will consider spacetimes, such as the Schwarzschild or the Reissner-N\"ordstrom, where is always posible to chose a tetrad so that the non zero spinor coefficients are real.
Finally, we will consider stress energy tensors of the form of the Reissner-Nordstr\"om one, so that the only non-zero Ricci scalar is $\Phi_{11}$. 

By performing a first order perturbation in Eq. \eqref{R321dB0_0} and Eq. \eqref{R321hB0_0}
%
one gets the following two equations:
\begin{eqnarray}
&{\bf D_{-1}}\,\Psi_{4}^{(1)} - {\bf \overline{\delta}_{-2}}\,\Psi_{3}^{(1)} + \left(3\,\eta\,\Psi_{2} + 2\Phi_{11}\right)\,\lambda^{(1)} 
- \varphi_1\,{{\bf \overline{\delta}}}_{-2} \,{\varphi_2}^{(1)}= 0,& \label{R321dB} \\
&- {\bf \delta}_{4}\,\Psi_{4}^{(1)} + {\bf \Delta}_{4}\,\Psi_{3}^{(1)} - \left(3\,\eta\,\Psi_{2} - 2\Phi_{11}\right)\,\nu^{(1)} 
- {\bf \Delta}_{2}\,\varphi_1\,{\varphi_2}^{(1)}=-\eta\,4\pi\,\,{\bf \overline{\delta}}_{0}\,{T_{22}}^{(1)}.&  \label{R321hB} 
\end{eqnarray}
%
In the forthcoming analysis
we consider that the external matter that causes the perturbation satisfies $T_{\mu\nu}^{(1)}\,k^\mu\,k^\nu \equiv T_{22}^{(1)}=\Phi_{22}^{(1)}/4 \pi \neq 0$. As we will show bellow, this condition is consistent with matter falling only in the radial direction.

In an analogous manner, the perturbation of Eq.~(\ref{eq:Psi4_0}) gives
\begin{equation}
\Psi _{4}^{(1)} + {\bf \Delta}_{2}\,{\lambda}^{(1)} - {\bf \overline{\delta}}_{-2}\,{\nu}^{(1)}=0 \ . \label{eq:BE4}    
\end{equation}
%
%
%

Finally, the following Ricci identities describing the action of the operators ${\bf D}$  and ${\bf \Delta}$ on the unperturbed fields also will be useful:
\begin{eqnarray}
D\,\Psi_2=\rho\,\left(3\eta\, \Psi_{2} + 2\Phi_{11}\right),  &\hspace{1cm}& D\,\Phi_{11}=4\,\eta\,\rho\,\Phi_{11}, \\
\Delta\,\Psi_2=-\mu\,\left(3\eta\,\Psi_{2} + 2\,\Phi_{11}\right),  &\hspace{1cm}& \Delta\,\Phi_{11}=-4\,\eta\,\mu\,\Phi_{11}, \\
D\,\varphi_1=2\,\eta\,\rho\,\varphi_1, &\hspace{1cm}& \Delta\,\varphi_1=-2\,\eta\,\mu \,\varphi_1. \label{eq:Del_phi1}
\end{eqnarray}
%
\subsection{System of equations for the perturbations}
\label{sec:System}
In the previous sub-section, we obtained four equations relating the perturbations  
$\Psi_{4}^{(1)}, \Psi_{3}^{(1)}, \varphi_2^{(1)}$, $\nu^{(1)}$ and
$\lambda^{(1)}$ due to the perturbed sources ${T_{22}}^{(1)}$ and $J_2^{(1)}$. It is an under-determined system, with four equations for five unknowns. However, as we will show, one can 
partially solve such system 
using the particular combination of  $\Psi_{3}^{(1)}$ and  $\varphi_2^{(1)}$ given by Eq. \eqref{eq:chi} 
and obtain a sub-system of coupled equations for $\Psi_4^{(1)}$ and $\chi$.
This remarkable combination has been related to a freedom in the rotation of the tetrad \cite{Chandrasekhar83}, although its physical meaning is not clearly understood and, to our knowledge, the physical meaning of such combination has not been discussed in the literature. In this section,  we present a detailed derivation
of such sub system of equations.

Acting with ${\bf \Delta}_{5}$ on Eq.~(\ref{R321dB}), and with ${\bf \overline{\delta}}_{-2}$ on Eq.~(\ref{R321hB}), adding and using the identity 
${\bf \Delta}_{1-q}\,{\bf \overline{\delta}}_{p} \equiv \,{\bf \overline{\delta}}_{p}\,{\bf \Delta}_{-q}$  with $q=-4, \,p=-2$
one gets
\begin{eqnarray}
&\left({\bf \Delta}_{5}\,{\bf D}_{-1} - {\bf \overline{\delta}}_{-2}\,{\bf \delta}_{4}\right)\,\Psi_{4}^{(1)} 
+ \left(\left(3\eta\,\Psi_{2} + 2\,\Phi_{11}\right)\,{\bf \Delta}_{5} + 3\,\eta\,\left(\Delta\,\Psi_2\right) 
+ 2\,\left(\Delta\,\Phi_{11}\right)\right)\,\lambda^{(1)}& \label{eq:paso0} \\
&-\left(3\eta\,\Psi_{2} - 2\,\Phi_{11}\right)\,{\bf \overline{\delta}}_{-2}\,\nu^{(1)}
- \varphi_1\,\left(\left({\bf \Delta}_{5} + \frac{\Delta\,\varphi_1}{\varphi_1}\right)\,{\bf \overline{\delta}}_{-2}
+ {\bf \overline{\delta}}_{-2}\,\left({\bf \Delta}_{2} + \frac{\Delta\,\varphi_1}{\varphi_1}\right)\right)\,{\varphi_2}^{(1)}=
-4\pi\,\eta\,{\bf \overline{\delta}}_{-2}\,{\bf \overline{\delta}}_{0}\,{T_{22}}^{(1)}. \nonumber 
\end{eqnarray}
A further simplification can be done
using the fact
that the action of the operator ${\bf \Delta}_N$ on $\varphi_1$, is
\begin{equation}
{\bf \Delta}_{N}\,(\varphi_1\,f )\equiv \varphi_1\,{\bf \Delta}_{N-2}\,f, \label{eq:comm}    
\end{equation}
for an arbitrary function $f$. 
Thus, using Eq. \eqref{eq:comm}, Eq. (\ref{eq:paso0}) takes the form
\begin{eqnarray}
&\left({\bf \Delta}_{5}\,{\bf D}_{-1} - {\bf \overline{\delta}}_{-2}\,{\bf \delta}_{4}\right)\,\Psi_{4}^{(1)} 
+ 3\eta\,\Psi_{2}\,\left(\left({\bf \Delta}_{2} + 3\,\eta\,\mu\right)\,\lambda^{(1)} - {\bf \overline{\delta}}_{-2}\,\nu^{(1)}\right) + 
2\,\Phi_{11}\left(\left({\bf \Delta}_{2} + 3\,\eta\,\mu\right)\,\lambda^{(1)} + {\bf \overline{\delta}}_{-2}\,\nu^{(1)}\right)
& \nonumber \\
&+ \left(3\,\eta\,(-\mu)\,\left(3\eta\,\Psi _{2} + 2\Phi_{11}\right) + 2\,\left(-4\,\eta\,\mu\,\Phi_{11}\right)\right)\,\lambda^{(1)}
- \varphi_1\,\left({\bf \Delta}_{3}\,{\bf \overline{\delta}}_{-2}
+ {\bf \overline{\delta}}_{-2}\,{\bf \Delta}_{0}\right)\,{\varphi_2}^{(1)}=
-4\pi\,\eta\,{\bf \overline{\delta}}_{-2}\,{\bf \overline{\delta}}_{0}\,{T_{22}}^{(1)}. \nonumber \\ \label{eq:paso1}
\end{eqnarray}
Substituting in the previous equation ${\bf \Delta}_{2}\,\lambda^{(1)}$
given by Eq.~(\ref{eq:BE4}) and after collecting terms 
%
one gets
\begin{eqnarray}
 &&\left({\bf \Delta}_{5}\,{\bf D}_{-1} - {\bf \overline{\delta}}_{-2}\,{\bf \delta}_{4} - 3\eta\,\Psi_{2} - 2\,\Phi_{11}\right)\,\Psi_{4}^{(1)} 
+ 4\,\Phi_{11}\,{\bf \overline{\delta}}_{-2}\,\nu^{(1)}   
\nonumber \\
&+& 
\mu\,\left(9\,\Psi_2 + 6\,\eta\,\Phi_{11} - 9\,\Psi_2 - 6\,\eta\,\Phi_{11} - 8\,\eta\,\Phi_{11}\right)\,\lambda^{(1)} - \varphi_1\,\left({\bf \Delta}_{3} + {\bf \Delta}_{1}\right)\,{\bf \overline{\delta}}_{-2}{\varphi_2}^{(1)}=
-4\pi\,\eta\,{\bf \overline{\delta}}_{-2}\,{\bf \overline{\delta}}_{0}\,{T_{22}}^{(1)}. \nonumber \\ 
\end{eqnarray}
Simplifying and using the definition of the operator ${\bf \Delta}_N$ in Eq.~(\ref{eq:OpsD_Del}), one obtains: 
%
\begin{eqnarray}
\left({\bf \Delta}_{5}\,{\bf D}_{-1} - {\bf \overline{\delta}}_{-2}\,{\bf \delta}_{4} - 3\,\eta\,\Psi_{2}- 2\,\Phi_{11}  \right)\,\Psi_{4}^{(1)}
- 2\,\varphi_1\,{\bf \Delta}_{2}\,{\bf \overline{\delta}}_{-2}{\varphi_2}^{(1)}
&-& 8\,\eta\,\mu\,\Phi_{11}\,\lambda^{(1)} + 4\,\Phi_{11}\,{\bf \overline{\delta}}_{-2}\,\nu^{(1)} \nonumber \\ 
&=&-4\pi\,\eta\,{\bf \overline{\delta}}_{-2}\,{\bf \overline{\delta}}_{0}\,{T_{22}}^{(1)}. \label{eq:paso2}
\end{eqnarray}
Next, one can use Eqs.~(\ref{R321dB}, \ref{R321hB}) 
to express the  perturbed spinors as:
\begin{eqnarray}
&\lambda^{(1)}=\frac{1}{2\,\Phi_{11} + 3\,\eta\,\Psi_{2}}\,\left(-{\bf D}_{-1}\,\Psi_{4}^{(1)} + {\bf \overline{\delta}}_{-2}\,\Psi_{3}^{(1)} 
+ \varphi_1\,{\bf \overline{\delta}}_{-2}\,{\varphi_2}^{(1)}\right),& \label{eq:lambda1} \\
&\nu^{(1)}=\frac{1}{2\Phi_{11} - 3\,\eta\,\Psi_{2}}\,\left({\bf \delta}_{4}\,\Psi_{4}^{(1)} 
- {\bf \Delta}_{4}\,\Psi_{3}^{(1)} + {\bf \Delta}_{2}\,\varphi_1\,{\varphi_2}^{(1)} - 4\pi\,\eta\,{\bf \overline{\delta}}_{0}\,{T_{22}}^{(1)}\right),&  
\label{eq:nu1}
\end{eqnarray}
and substitute  them 
in Eq.~(\ref{eq:paso2}). The resulting equation is:
\begin{equation}
{\bf A}_1\,\Psi_{4}^{(1)} + {\bf A}_2\,{\varphi_2}^{(1)} + {\bf A}_3\,\Psi_{3}^{(1)}=4\pi\,{\bf A}_4\,{T_{22}}^{(1)},  \label{eq:paso4}
\end{equation}
where we have defined the operators
\begin{eqnarray}
{\bf A}_{1}
&=&\left(\Delta + 3\,\eta\,\mu\,\frac{6\,\Phi_{11} + 5\,\eta\,\Psi_2}{(2\,\Phi_{11} + 3\,\eta\,\Psi_2)^2}\right)\,{\bf D}_{-1} 
- 2\,\Phi_{11} - 3\eta\,\Psi_{2} + \frac{2\,\Phi_{11} + 3\,\eta\,\Psi_{2}}{2\,\Phi_{11} - 3\eta\, \Psi_{2}}\,
{\bf \overline{\delta}}_{-2}\,{\bf \delta}_{4}, \\ 
{\bf A}_{2}
&=& \frac{\left(2\,\varphi_1\right)\,\left(3\eta\,\Psi_2\right)}{2\,\Phi_{11} - 3\,\eta\,\Psi_2}\,
\left(\Delta - 
\frac{2\,\eta\,\mu\left(2\Phi_{11}^2 - 3\,\Psi_2^2 - 3\,\Phi_{11}\,\eta\,\Psi_2\right)}{\eta\,\Psi_2\,\left(2\,\Phi_{11} + 
3\,\eta\,\Psi_{2}\right)}\right)\,{\bf \overline{\delta}}_{-2}, \label{op:A2}\\
{\bf A}_{3}
&=&-\frac{4\,\Phi_{11}}{2\,\Phi_{11} - 3\,\eta\,\Psi_{2}}\,\left(\Delta + 
\eta\,\mu\,\frac{14\,\Phi_{11}+ 9\,\eta\,\Psi_2}{2\,\Phi_{11} + 3\,\eta\,\Psi_2}\right)\,{\bf \overline{\delta}}_{-2}, \label{op:A3}\\
{\bf A}_{4}
&=&\eta\,\frac{2\,\Phi_{11} + 3\,\eta\,\Psi_{2}}{2\,\Phi_{11} - 3\,\eta\,\Psi_{2}}\,{\bf \overline{\delta}}_{-2}\,
{\bf \overline{\delta}}_{0}.
\end{eqnarray}
Remarkably 
the operator ${\bf A}_{2}$, acting on $-3\,\eta\,\Psi_2\,\varphi_{2}^{(1)}$  takes the same form as the operator ${\bf A}_{3}$ acting on $2\,\eta\,\varphi_1\,\Psi_3^{(1)}$. Indeed
one can writte the action of  ${\bf A}_{2}$ and  ${\bf A}_{3}$ on $\varphi_2^{(1)}$ and
$\Psi_3^{(1)}$ as
%
\begin{eqnarray}
{\bf A}_{2}\,\varphi_2^{(1)}&=&{\bf A}_{2}\,\left(\frac{1}{(-3\,\eta\,\Psi_2)}\,(-3\,\eta\,\Psi_2\,\varphi_2^{(1)})\right)\nonumber \\
&=&-\frac{\left(2\,\varphi_1\right)}{\left(2\,\Phi_{11} - 3\,\eta\,\Psi_2\right)}\,
\left(\Delta + 3\,\eta\,\mu\,\frac{6\,\Phi_{11} + 5\,\eta\,\Psi_{2}}{2\,\Phi_{11} + 
3\,\eta\,\Psi_{2}}\right)\,{\bf \overline{\delta}}_{-2}\,\left(-3\,\eta\,\Psi_2\,\varphi_2^{(1)}\right), \\
{\bf A}_{3}\,\Psi_3^{(1)}&=&{\bf A}_{3a}\,
\left(\frac{1}{2\,\eta\,\varphi_1}\,(2\,\eta\,\varphi_1\,\Psi_3^{(1)})\right) \nonumber \\
&=&-\frac{\left(2\,\varphi_1\right)}{\left(2\,\Phi_{11} - 3\,\eta\,\Psi_2\right)}\,
\left(\Delta + 3\,\eta\,\mu\,\frac{6\,\Phi_{11} + 5\,\eta\,\Psi_{2}}{2\,\Phi_{11} + 
3\,\eta\,\Psi_{2}}\right)\,{\bf \overline{\delta}}_{-2}\,\left(2\,\eta\,\varphi_1\,\Psi_3^{(1)}\right),
\end{eqnarray}
and using this remarkable property in Eq.~(\ref{eq:paso4}), one gets:
\begin{equation}
{\bf A}_{1}\,\Psi_{4}^{(1)} + {\bf A}_{5}\,\chi =4\pi\,{\bf A}_{4}\,{T_{22}}^{(1)}, \label{eq:sys1f}
\end{equation}
%
%
where
\begin{equation}
\chi =\eta\,\left(2\,\varphi_1\,\Psi_3^{(1)} - 3\,\Psi_2\,\varphi_2^{(1)}\right), \label{def:chi}    
\end{equation}
%
%
and
\begin{equation}
{\bf A}_{5}=-\frac{\left(2\,\varphi_1\right)}{\left(2\,\Phi_{11} - 3\,\eta\,\Psi_2\right)}\,
\left(\Delta + 3\,\eta\,\mu\,\frac{6\,\Phi_{11} + 5\,\eta\,\Psi_{2}}{2\,\Phi_{11} + 
3\,\eta\,\Psi_{2}}\right)\,{\bf \overline{\delta}}_{-2}.
\end{equation}

In order to derive a second equation for $\Psi_4^{(1)}$, and $\chi$
we first apply the operator ${\bf \delta}_{4}$ on Eq.~(\ref{R321dB}), and the operator ${\bf D}_{-2}$ on Eq.~(\ref{R321hB}),
adding up  and using the rule of commutation ${\bf \delta}_{4}\,{\bf D}_{-1} -
{\bf D}_{-2}\,{\bf \delta}_{4}=0$, 
we eliminate $\Psi_4^{(1)}$ in the equation. The resulting equation is
\begin{eqnarray}
&&\left({\bf D}_{-2}\,{\bf \Delta}_{4} - {\bf \delta}_{4}\,{\bf \overline{\delta}}_{-2}\right)\Psi_{3}^{(1)} +
\left(3\eta\, \Psi _{2} + 2\Phi _{11}\right)\,{\bf \delta}_{4}\,\lambda^{(1)} 
 +
\left(-\varphi_1\,{\bf \delta}_{4}\,{\bf \overline{\delta}}_{-2} - {\bf D}_{-2}\,{\bf \Delta}_{2}\,\varphi_1\right)\,{\varphi_2}^{(1)}-
\nonumber \\ 
&&
\left[\left(3\eta\,\Psi _{2} - 2\,\Phi _{11}\right)\,{\bf D}_{-2} + 3\eta\,\left(D\,\Psi _{2}\right) - 
2\,\left(D\,\Phi _{11}\right)\right]\,\nu^{(1)}=-4\pi\,\eta\,{\bf D}_{-2}\,{\bf \overline{\delta}}_{0}\,{T_{22}}^{(1)}.
\end{eqnarray}
Considering that ${\bf D}_{-2}={\bf D}_{+1} - 3\,\eta\,\rho_s$ we
can use the Maxwell equation Eq.~(\ref{eq:maxphi3c})
to expand this last equation as:
\begin{eqnarray}
&&\left({\bf D}_{-2}\,{\bf \Delta}_{4} - {\bf \delta}_{4}\,{\bf \overline{\delta}}_{-2}\right)\Psi_{3}^{(1)} 
- \varphi_1\,\left(\left({\bf D}_{-2} + 2\,\eta\,\rho\right)\,\left({\bf \Delta}_{2} - 2\,\eta\,\mu\right) + 
{\bf \delta}_{4}\,{\bf \overline{\delta}}_{-2}\right)\,{\varphi_2}^{(1)} \nonumber \\ 
&&- 3\eta\, \Psi _{2}\,\left(\frac{1}{2\,\eta\,\varphi_1}\,\left[{\bf D}_{-1}\,{\bf \Delta}_{3} - 
{\bf \delta}_{4}\,{\bf \overline{\delta}}_{-2} - 6\,\eta\,\Psi_2\right]\,{\varphi_2}^{(1)} + 
2\,\Psi_{3}^{(1)} + \frac{1}{4\,\varphi_1}
\,J_2^{(1)} \right) +  \nonumber \\  
&& 2\Phi _{11}\,\left(2\,{\bf \delta}_{4}\,\lambda^{(1)} + \frac{1}{2\,\eta\,\varphi_1}\,\left[{\bf D}_{-1}\,{\bf \Delta}_{3} - 
{\bf \delta}_{4}\,{\bf \overline{\delta}}_{-2} - 6\,\eta\,\Psi_2\right]\,{\varphi_2}^{(1)} +
2\,\Psi_{3}^{(1)} + \frac{1}{4\,\varphi_1}\,J_2^{(1)}\right)  - 4\, \eta \,\rho  \,\Phi _{11}\,\nu^{(1)}
\nonumber \\   
&& =-4\pi\,\eta\,{\bf D}_{-2}\,{\bf \overline{\delta}}_{0}\,{T_{22}}^{(1)}. 
\end{eqnarray}
After some algebra the previous equation becomes
\begin{eqnarray}
&&\left({\bf D}_{-2}\,{\bf \Delta}_{4} - {\bf \delta}_{4}\,{\bf \overline{\delta}}_{-2} - 
2\,\left(3\eta\,\Psi _{2} - 2\,\Phi _{11}\right)\right)\Psi_{3}^{(1)} +
\nonumber \\&&- \left(\varphi_1\,\left({\bf D}_{0}\,{\bf \Delta}_{0} + {\bf \delta}_{4}\,{\bf \overline{\delta}}_{-2}\right)
+ \frac{3\eta\,\Psi _{2} - 2\,\Phi _{11}}{2\,\eta\,\varphi_1}\left( {\bf D}_{-1}\,{\bf \Delta}_{3} - 
{\bf \delta}_{4}\,{\bf \overline{\delta}}_{-2} - 6\,\eta\,\Psi_2 \right)\right)\,{\varphi_2}^{(1)} + \nonumber \\   
&& 4\,\Phi _{11}\,{\bf \delta}_{4}\,\lambda^{(1)} - 4 \, \eta \,\rho  \,\Phi _{11}\,\nu^{(1)}
=-4\pi\,\eta\,{\bf D}_{-2}\,{\bf \overline{\delta}}_{0}\,{T_{22}}^{(1)} + 
\frac{1}{4\,\varphi_1}\,\left(3\eta\,\Psi _{2} - 2\,\Phi _{11}\right)\,J_2^{(1)}.  \label{eq:psi3}
\end{eqnarray}
From Eqs. (\ref{R321dB}, \ref{R321hB}) one can obtain the following expressions for the terms involving the perturbed spinor coefficients:
\begin{eqnarray}
4\,\Phi _{11}\,{\bf \delta}_{4}\,\lambda^{(1)}&=&\frac{4\,\Phi _{11}}{2\,\Phi _{11} + 3\,\eta\,\Psi_2}\,
\left(- {\bf D}_{-2}\,{\bf \delta}_{4}\,\Psi_4^{(1)} + {\bf \delta}_{4}\,{\bf \overline{\delta}}_{-2}\,\Psi_{3}^{(1)}
+ \varphi_1\,{\bf \delta}_{4}\,{\bf \overline{\delta}}_{-2}\,\varphi_2^{(1)}\right), \label{eq:dlam} \\ 
-4\,\eta\,\rho\,\Phi _{11}\,\nu^{(1)}&=&-\eta\,\rho\frac{4\,\Phi _{11}}{2\,\Phi _{11} - 3\,\eta\,\Psi_2}\,
\left({\bf \delta}_{4}\,\Psi_4^{(1)} - {\bf \Delta}_4\,\Psi_{3}^{(1)}
+ \varphi_1\,{\bf \Delta}_{0}\,\varphi_2^{(1)} - 4\pi\,\eta\,{\bf \overline{\delta}}_{0}\,T_{22}^{(1)}\right). \label{eq:r-nu}
\end{eqnarray}
Using this expressions, 
in Eq. (\ref{eq:psi3}), and after some simplifications we obtain an equation involving only $\Psi_4^{(1)}, \Psi_3^{(1)}, \varphi_2^{(1)}$, and the sources:
\begin{equation}
{\bf B}_1 \, \Psi_4^{(1)} + \left({\bf O}_{1r} + {\bf O}_{1a}\right)\,\Psi_3^{(1)} + \left({\bf O}_{2r} + {\bf O}_{2a}\right)\,\varphi_2^{(1)} 
=4\pi\,{\bf B}_3\,{T_{22}}^{(1)} + \frac{1}{4\,\varphi_1}\,\left(3\eta\,\Psi _{2} - 2\,\Phi _{11}\right)\,J_2^{(1)}, \label{eq:system1}
\end{equation}
where the operators have the form:
\begin{align}
{\bf B}_{1}
&=-\frac{4\,\Phi_{11}}{2\,\Phi_{11} + 3\,\eta\,\Psi_2}\,\left({\bf D}_{0} -  
\eta \,\rho \, \frac{2\,\Phi_{11} - 9\,\eta\,\Psi_2}{2\,\Phi_{11} - 3\,\eta\,\Psi_2}\right)\,{\bf \delta}_{4}, \label{Op:B1}\\
{\bf O}_{1r}
&=\left({\bf D}_{0} + 6\,\eta\,\rho\,\frac{\eta\,\Psi_{2}}{2\,\Phi_{11} - 3\,\eta\,\Psi_2}\right)\,\left({\bf \Delta}_0 + 4\,\eta\,\mu\right), 
\label{Op_1r}\\
{\bf O}_{1a}&=\frac{2\,\Phi_{11} - 3\,\eta\,\Psi_2}{2\,\Phi_{11} + 3\,\eta\,\Psi_2}\,
\left[{\bf \delta}_{4}\,{\bf \overline{\delta}}_{-2}  + 2\,\left(2\,\Phi_{11} + 3\,\eta\,\Psi_2\right)\right], \label{Op_1a} \\
{\bf O}_{2r}
&=-\varphi_1\,\left({\bf D}_{0} + 4\,\eta\,\rho\,\frac{\Phi_{11}}{2\,\Phi_{11} - 3\,\eta\,\Psi_2}\right)\,{\bf \Delta}_{0}
+ \frac{2\,\Phi_{11} - 3\,\eta\,\Psi_2}{2\,\eta\,\varphi_1}\,\left({\bf D}_{0} - \eta\,\rho\right)\,\left({\bf \Delta}_{0} + 3\,\eta\,\mu\right), 
\label{Op_2r}\\
{\bf O}_{2a}&=-\left(\frac{3\,\eta\,\Psi_2}{2\,\eta\,\varphi_1}\right)\,\frac{2\,\Phi_{11} - 3\,\eta\,\Psi_2}{2\,\Phi_{11} + 3\,\eta\,\Psi_2}\,
\left[{\bf \delta}_{4}\,{\bf \overline{\delta}}_{-2}  + 2\,\left(2\,\Phi_{11} + 3\,\eta\,\Psi_2\right)\right], \label{Op_2a}\\
{\bf B}_{3}
&=-\eta\,\left({\bf D}_{0} + 6\,\eta\,\rho\,\frac{\eta\,\Psi_{2}}{2\,\Phi_{11} - 3\,\eta\,\Psi_2}\right)\,\overline \delta_0. \label{Op_4}
\end{align}
We have collected the operators acting on $\Psi_3^{(1)}$ and on $\varphi_2^{(1)}$ in those involving ${\bf D}_N$ and ${\bf \Delta}_N$, and the rest, as long as they involve more algebraic manipulation in the next steps in the derivation.
Indeed, 
using $\Psi_3^{(1)}=
(2\eta\varphi_1/2\eta\varphi_1)\,\Psi_3^{(1)}$ 
and $\varphi_2^{(1)}=(-3\eta\Psi_2/-3\eta\Psi_2)\,\varphi_2^{(1)}$ in ${\bf O}_{1a}\Psi_3^{(1)}$ one gets 
\begin{equation}
{\bf O}_{1a}
\Psi_3^{(1)}  =
{\bf O}_{1a}\left(
\frac{2\eta\varphi_1}{2\eta\varphi_1}\,\Psi_3^{(1)} \right) = {\cal O}_{1a}(2\eta\varphi_1 \Psi_3^{(1)}), 
\end{equation}
%
%
and for the radial operators
\begin{eqnarray}
{\bf O}_{1r}\,\Psi_3^{(1)}&=&{\bf O}_{1r}\,\frac{1}{(2\,\eta\,\varphi_1)}\,\left(2\,\eta\,\varphi_1\,\Psi_3^{(1)}\right)
={\cal O}_{1r}\,\left(2\,\eta\,\varphi_1\,\Psi_3^{(1)}\right),\\
{\bf O}_{2r}\,\varphi_2^{(1)}&=&{\bf O}_{2r}\,\frac{1}{(-3\,\eta\,\Psi_2)}\left(-3\,\eta\,\Psi_2\,\varphi_2^{(1)}\right)=
{\cal O}_{2r}\,\left(-3\,\eta\,\Psi_2\,\varphi_2^{(1)}\right),
\end{eqnarray}
where ${\cal O}_{1a}$, ${\cal O}_{1r}$, and ${\cal O}_{2r}$ have the form:
\begin{eqnarray}
{\cal O}_{1a}=\left(\frac{1}{2\,\eta\,\varphi_1}\right)\,\frac{2\,\Phi_{11} - 3\,\eta\,\Psi_2}{2\,\Phi_{11} + 3\,\eta\,\Psi_2}\,
\left({\bf \delta}_{4}\,{\bf \overline{\delta}}_{-2}  + 2\,\left(3\,\eta\,\Psi_2 + 2\,\Phi_{11}\right)\right),
\end{eqnarray}
and
\begin{eqnarray}
{\cal O}_{1r}
&=&\frac{1}{2\,\eta\,\varphi_1}\,\left({\bf D}_{-2} - \frac{D\,\varphi_1}{\varphi_1} + 
4 \, \eta \,\rho\,\frac{\Phi_{11}}{2\,\Phi_{11} - 3\,\eta\,\Psi_2}\right)\,\left({\bf \Delta}_4 - \frac{\Delta\,\varphi_1}{\varphi_1}\right)
\label{eq:OpO1r} \\
&=& \frac{1}{2\,\eta\,\varphi_1}\,\left({\bf D}_{0}\,{\bf \Delta}_0 + 6\,\eta\,\mu\,{\bf D}_{0} - 
4\,\eta\,\rho\,\frac{\Phi_{11} - 3\,\eta\,\Psi_2}{2\,\Phi_{11} - 3\,\eta\,\Psi_2}\,{\bf \Delta}_0 + 6\,\eta\,\left(D\,\mu\right)
- 24\,\rho\,\mu\,\frac{\Phi_{11} - 3\,\eta\,\Psi_2}{2\,\Phi_{11} - 3\,\eta\,\Psi_2}\right),  \nonumber \\
{\cal O}_{2r}&=&-\frac{1}{3\,\eta\,\Psi_2}\left[-\varphi_1\,\left({\bf D}_{0} - \frac{D\,\Psi_2}{\Psi_2} + 
4 \, \eta \,\rho\,\frac{\Phi_{11}}{2\,\Phi_{11} - 3\,\eta\,\Psi_2}\right)\,\left({\bf \Delta}_{0} - \frac{\Delta\,\Psi_2}{\Psi_2}\right)
\right. \nonumber \\
&&  \left.
+ \frac{2\,\Phi_{11} - 3\,\eta\,\Psi_2}{2\,\eta\,\varphi_1}\,\left({\bf D}_{-1} - \frac{D\,\Psi_2}{\Psi_2}\right)\,
\left({\bf \Delta}_{3} - \frac{\Delta\,\Psi_2}{\Psi_2}\right)\right] \nonumber \\
&=&a_0\,{\bf D}_{0}\,{\bf \Delta}_0 + a_1\,{\bf D}_{0} + a_2\,{\bf \Delta}_0 + a_3\,\left(D\,\mu\right) + a_4. \label{eq:OpO2r}
\end{eqnarray}
We have written the operator ${\cal O}_{2r}$ as a sum of different operators, since as shown below each element of the sum is equal to the corresponding element of ${\cal O}_{1r}$:
\begin{eqnarray}
a_0&=&-\frac{1}{3\,\eta\,\Psi_2}\,\left(-\varphi_1 + \frac{2\,\Phi_{11} - 3\,\eta\,\Psi_2}{2\,\eta\,\varphi_1}\right)=\frac{1}{2\,\eta\,\varphi_1}, \\
a_1&=&-\frac{1}{3\,\eta\,\Psi_2}\,\frac{\eta\,\mu}{2\,\eta\,\varphi_1\,\eta\,\Psi_2}\,
\left(-2\,\Phi_{11}\,\left(2\,\Phi_{11} + 3\,\eta\,\Psi_2\right) + 2\,\left(2\,\Phi_{11} - 3\,\eta\,\Psi_2\right)\,
\left(\Phi_{11} + 3\,\eta\,\Psi_2\right)\right)=\frac{1}{2\,\eta\,\varphi_1}\,6\,\eta\,\mu, \\
a_2&=&-\frac{1}{3\,\eta\,\Psi_2}\,\frac{-\eta\,\rho}{2\,\eta\,\varphi_1\,\eta\,\Psi_2\,\left(2\,\Phi_{11} - 3\,\eta\,\Psi_2\right)}\,
(-2\,\Phi_{11}\,\left(4\,\Phi_{11}^2 -4\,\eta\,\Psi_2\,\Phi_{11} -9\,\eta\,\Psi_2^2\right) 
\nonumber \\&+&
2\,\left(2\,\Phi_{11} - 3\,\eta\,\Psi_2\right)^2\,
\left(\Phi_{11} + 2\,\eta\,\Psi_2\right)) 
=\frac{1}{2\,\eta\,\varphi_1}\,\left(-4\,\eta\,\rho\frac{\Phi_{11} - 3\,\eta\,\Psi_2}{2\,\Phi_{11} - 3\,\eta\,\Psi_2}\right), \\
a_3&=&-\frac{1}{3\,\eta\,\Psi_2}\,\frac{\eta}{\eta\,\Psi_2\,2\,\eta\,\varphi_1}\left(-2\,\Phi_{11}\,\left(2\,\Phi_{11} + 3\,\eta\,\Psi_2\right)
+ \left(2\,\Phi_{11} - 3\,\eta\,\Psi_2\right)\,\left(\Phi_{11} + 3\,\eta\,\Psi_2\right)\right)=\frac{1}{2\,\eta\,\varphi_1}\,6\,\eta, \\
a_4&=&-\frac{1}{3\,\eta\,\Psi_2}\,\left(-\varphi_1\,\left(2\,\mu\,\left(D\frac{\Phi_{11}}{\Psi_2}\right) - \rho\,\mu\,
\frac{\left(4\,\Phi_{11}^2 -4\,\eta\,\Psi_2\,\Phi_{11} -9\,\Psi_2^2\right)\,
\left(2\,\Phi_{11} + 3\,\eta\,\Psi_2\right)}{\Psi_2^2\,\left(2\,\Phi_{11} - 3\,\eta\,\Psi_2\right)}\right) \right. \nonumber \\
&& \left. + \frac{2\,\Phi_{11} - 3\,\eta\,\Psi_2}{2\,\eta\,\varphi_1}\,\left(2\,\mu\,\left(D\frac{\Phi_{11}}{\Psi_2}\right) 
- 4\,\rho\,\mu\,\frac{\left(\Phi_{11} + 2\,\eta\,\Psi_2\right)\,\left(\Phi_{11} + 3\,\eta\,\Psi_2\right)}{\Psi_2^2}\right)\right) \nonumber \\
&=&\frac{1}{2\,\eta\,\varphi_1}\,\left(-24\,\rho\,\mu\,\frac{\Phi_{11} - 3\,\eta\,\Psi_2}{2\,\Phi_{11} - 3\,\eta\,\Psi_2}\right),
\end{eqnarray}
thus,
each of the coefficients of ${\cal O}_{2r}$ take the same form as the corresponding coefficients of ${\cal O}_{1r}$
acting on the variable $\chi$ defined in Eq.~(\ref{def:chi}).
As a result, we can again express the operators acting on $\Psi_3^{(1)}$ with the 
operators acting on $\varphi_2^{(1)}$, as a single operator acting on $\chi$, and obtain in this way the second equation for $\Psi_4^{(1)}$ and $\chi$
%
\begin{equation}
{\bf B}_1\,\Psi_4^{(1)} + {\bf B}_1\,\chi  ={\bf B}_3\,\kappa\,{T_{22}}^{(1)} + 
\frac{1}{4\,\varphi_1}\,\left(3\eta\,\Psi _{2} - 2\,\Phi _{11}\right)\,J_2^{(1)}, \label{eq:sys2f}
\end{equation}
with ${\bf B}_1 ={\cal O}_{1r} + {\cal O}_{1a}$. 
We have shown that it is possible to get a couple of equations for $\Psi_4^{(1)}$ and $\chi$ that are independent of $\lambda^{(1)}$ and $\nu^{(1)}$. However, the complete system is not solved as long as, in order to determine the perturbed spinor coefficients one should obtain 
$\Psi_4^{(1)}$ and $\Psi_3^{(1)}$ and $\varphi_2^{(1)}$ independently, which is not possible within this formalism because, as mentioned above, the system is under determined.

\section{Sources of the perturbation and harmonic decomposition}
\label{sec:sources}

Since the previous section appears quite lengthy due to all the calculations, we find helpful to present in short the two important equations that describe the gravitational and electromagnetic/gravitational perturbations for $\Psi_4^{(1)}$ and $\chi$ 
and introduce further comments on the sources that may cause these perturbations. From Eqs. (\ref{eq:system1}, \ref{eq:sys2f}), we have:
%
\begin{eqnarray}
A_{1}\,\Psi_{4}^{(1)} + A_{5}\,\chi &=&A_{4}\,\kappa\,{T_{22}}^{(1)}, \label{eq:d2Psi4}\\
B_{1}\,\Psi_{4}^{(1)} + B_{2}\,\chi &=&B_{3}\,\kappa\,{T_{22}}^{(1)} + \frac{1}{4\,\varphi_1}\,\left(2\,\Phi_{11} - 3\,\eta\,\Psi_{2}\right)\,J_2^{(1)}, 
\label{eq:d2chi}
\end{eqnarray}
where
\begin{eqnarray}
\chi &=&\eta\,\left(2\,\varphi_1\,\Psi_3^{(1)} - 3\,\Psi_2\,\varphi_2^{(1)}\right), \\
A_1&=&\left(\Delta + 3\,\eta\,\mu\,\frac{6\,\Phi_{11} + 5\,\eta\,\Psi_2}{2\,\Phi_{11} + 3\,\eta\,\Psi_2}\right)\,\left({\bf D}_{0} - \eta\,\rho\right) 
- 2\,\Phi_{11} - 3\eta\,\Psi_{2} + \frac{2\,\Phi_{11} + 3\,\eta\,\Psi_{2}}{2\,\Phi_{11} - 3\eta\, \Psi_{2}}\,
{\bf \overline{\delta}}_{-2}\,{\bf \delta}_{4},  \\ 
A_{5}&=&-\frac{\left(2\,\varphi_1\right)}{\left(2\,\Phi_{11} - 3\,\eta\,\Psi_2\right)}\,
\left(\Delta + 3\,\eta\,\mu\,\frac{6\,\Phi_{11} + 5\,\eta\,\Psi_{2}}{2\,\Phi_{11} + 
3\,\eta\,\Psi_{2}}\right)\,{\bf \overline{\delta}}_{-2},  \\
A_4&=&\eta\,\frac{2\,\Phi_{11} + 3\,\eta\,\Psi_{2}}{2\,\Phi_{11} - 3\,\eta\,\Psi_{2}}\,{\bf \overline{\delta}}_{-2}\,
{\bf \overline{\delta}}_{0}, \\
B_1&=&-\frac{4\,\Phi_{11}}{2\,\Phi_{11} + 3\,\eta\,\Psi_2}\,\left({\bf D}_{0} -  
\eta \,\rho \, \frac{2\,\Phi_{11} - 9\,\eta\,\Psi_2}{2\,\Phi_{11} - 3\,\eta\,\Psi_2}\right)\,{\bf \delta}_{4}, \\
B_2&=&\frac{1}{2\,\eta\,\varphi_1}\,\left(\left({\bf D}_{0} - 4\,\eta\,\rho\,\frac{\Phi_{11} - 3\,\eta\,\Psi_2}{2\,\Phi_{11} - 3\,\eta\,\Psi_2}\right)\,
\left({\bf \Delta}_0 + 6\,\eta\,\mu\right) + 2\,\left(2\,\Phi_{11} - 3\eta\,\Psi_{2}\right) + 
\frac{2\,\Phi_{11} - 3\,\eta\,\Psi_{2}}{2\,\Phi_{11} + 3\eta\, \Psi_{2}}\,
{\bf \delta}_{4}\,{\bf \overline{\delta}}_{-2}\right), \\
B_3&=&-\eta\,\left({\bf D}_{0} + 6\,\eta\,\rho\,\frac{\eta\,\Psi_{2}}{2\,\Phi_{11} - 3\,\eta\,\Psi_2}\right)\,\overline \delta_0.
\end{eqnarray} 
%
As the source of perturbation, let us consider a charged dust like matter falling radially into the black hole 
with stress energy tensor ${T_{\mu\nu}}=\rho_f u^{\mu} u^{\nu}$, 
where $u^{\mu}$ is the four velocity and $\rho_f$ is the rest mass density of the fluid.
In our analysis we consider that the fluid is
falling radially into the black hole with four velocity:
\begin{eqnarray}
u^{\mu}= (u^t
(t, r), u^r
(t, r), 0, 0) \ .
\end{eqnarray}
When the fluid is charged it induces an electric current given by
${J^\mu}^{(1)}=\rho_e\,{u_e}^\mu$ where the electric density is $\rho_e = q \rho_f$, and $q$ is the charge per unit of mass of each particle. 

An important property of the system of Eqs. (\ref{eq:d2Psi4}, \ref{eq:d2chi}) with the given form of the sources, is that the system can be decoupled into an angular and radial set of equations.

First, one must notice that the different functions of the system Eqs.~(\ref{eq:d2Psi4}, \ref{eq:d2chi}), have different spin weight. For instance, the Weyl scalar $\Psi_4^{(1)}$ has spin weight minus two, whereas $\chi$ has spin weight minus one, also the rest mass density $\rho_n$,  and the density of charge $\rho_e$, are scalar functions with zero spin. Thus, the different operators acting on these functions have to be such that they finally produce quantities with the same spin weight, indeed, as we show bellow, Eq.~(\ref{eq:d2Psi4}) has spin weight $-2$ and Eq.~(\ref{eq:d2chi}) spin weight $-1$. With this in mind and using the fact that the spin-weighted spherical harmonics ${Y_{s}}^{(l,m)}$ form a basis for each weight $s$, we write 
\begin{eqnarray}
{\Psi_4}^{(1)}&=&\sum_{l,m}\,{\Psi_4}_{l\,m}(t,r)\,{Y_{-2}}^{(l,m)}(\theta, \varphi),  \\
\chi&=&\sum_{l,m}\,\chi_{l\,m}(t,r)\,{Y_{-1}}^{(l,m)}(\theta, \varphi), 
\\
\rho_n&=&\sum_{l,m}\,{\rho_n}_{l\,m}(t,r)\,{Y_0}^{(l,m)}(\theta, \varphi),\label{eq:expansion_n} \\
\rho_e&=&\sum_{l,m}\,{\rho_e}_{l\,m}(t,r)\,{Y_0}^{(l,m)}(\theta, \varphi),\label{eq:expansion_e}
\end{eqnarray}
In spherical symmetry we can choose the vector $k^{\mu} = \left( 1,  -1, 0, 0\right)$ such that the
operators 
$\delta_N=m^\mu\,\partial_\mu$ and $\bar{\delta}_N$ in Eq.~(\ref{def:deltas}), can be written in terms of the eth and eth-bar operators, $\eth_s=-\left(\partial_\theta + i\,\csc\theta\,\partial_\varphi - s\,\cot\theta\right)$ 
and $\bar{\eth}_s=-\left(\partial_\theta - i\,\csc\theta\,\partial_\varphi + s\,\cot\theta\right)$  as:
\begin{equation}
\delta_N=-\frac1{\sqrt{2}\,r}\eth_{-\frac{N}2}, \qquad  \bar{\delta}_N=-\frac1{\sqrt{2}\,r}{\,\bar{\eth}}_{\frac{N}2}.
\end{equation}
The action of the eth, eth-bar operators on  ${Y_s}^{(l,\,m)}$ is to rise or lower the spin weight as: 
\begin{eqnarray}
\eth_s\,{Y_s}^{(l,m)}&=&\sqrt{\left(l-s\right)\,\left(l+s+1\right)}\,{Y_{s+1}}^{(l,m)}~, \label{op:eth_Y} \\
{\bar \eth}_s\,{Y_s}^{l,m}&=&-\sqrt{\left(l+s\right)\,\left(l-s+1\right)}\,{Y_{s-1}}^{(l,m)}~. \label{op:beth_Y}
\end{eqnarray}
Given the action of the operators on the spin-weight, one gets the following identities for the terms in the perturbation equations:
%
\begin{eqnarray}
&{\bf \overline{\delta}}_{-2}\,{\bf \delta}_{4}\,{\Psi_4}^{(1)}&=\frac{1}{2\,r^2}\,{\bar{\eth}}_{-1}\,\eth_{-2}\,
\sum_{l,m}\,{\Psi_4}_{l\,m}\,{Y_{-2}}^{(l,m)}=-\frac{1}{2\,r^2}\,\sum_{l,m}\,\left(l-1\right)\,\left(l+2\right)\,{\Psi_4}_{l\,m}\,{Y_{-2}}^{(l,m)}, 
\nonumber \\
&{\bf \overline{\delta}}_{-2}\,\chi&=-\frac{1}{\sqrt{2}\,r}\,{\bar{\eth}}_{-1}\,\sum_{l,m}\,\chi_{l\,m}\,{Y_{-1}}^{(l,m)}=
-\frac{1}{\sqrt{2}\,r}\,\sum_{l,m}\,\sqrt{\left(l-1\right)\,\left(l+2\right)}\,\chi_{l\,m}\,{Y_{-2}}^{(l,m)},\nonumber \\
&{\bf \overline{\delta}}_{-2}\,{\bf \overline{\delta}}_{0}\,\kappa\,{T_{22}}^{(1)}&=\kappa\,\frac{\left(k^\mu\,u_\mu\right)^2}{2\,r^2}
{\bar \eth}_{-1}\,{\bar \eth}_0\,\sum_{l,m}\,{\rho_n}_{l\,m}(t,r)\,{Y_0}^{(l,m)},
\nonumber \\&&=\kappa\,\frac{\left(k^\mu\,u_\mu\right)^2}{2\,r^2}\,
\sum_{l,m}\,\sqrt{\left(l-1\right)\,l\,\left(l+1\right)\,\left(l+2\right)}{\rho_n}_{l\,m}(t,r)\,{Y_{-2}}^{(l,m)}, \nonumber \\
&\delta_4\,{\Psi_4}^{(1)}&=-\frac{1}{\sqrt{2}\,r}\,{\bar{\eth}}_{-2}\sum_{l,m}\,\sum_{l,m}\,{\Psi_4}_{l\,m}\,{Y_{-2}}^{(l,m)}=
-\frac{1}{\sqrt{2}\,r}\,\sum_{l,m}\,\sqrt{\left(l-1\right)\,\left(l+2\right)}\,{\Psi_4}_{l\,m}\,{Y_{-1}}^{(l,m)}, \nonumber \\
&\delta_4\,{\bf \overline{\delta}}_{-2}\,\chi &=\frac{1}{2\,r^2}\,\eth_{-2}\,{\bar \eth}_{-1}\sum_{l,m}\,\chi_{l\,m}\,{Y_{-1}}^{(l,m)}= 
-\frac{1}{2\,r^2}\,\sum_{l,m}\,\left(l-1\right)\,\left(l+2\right)\,\chi_{l\,m}\,{Y_{-1}}^{(l,m)}, \nonumber \\
&\delta_0\,\kappa\,{T_{22}}^{(1)}&=-\kappa\,\frac{\left(k^\mu\,u_\mu\right)^2}{\sqrt{2}\,r}\,{\bar \eth}_0\,\sum_{l,m}\,{\rho_n}_{l\,m}\,{Y_0}^{(l,m)}=
\frac{\left(k^\mu\,u_\mu\right)^2}{\sqrt{2}\,r}\,\sum_{l,m}\,\sqrt{l\,\left(l+1\right)}\,{\rho_n}_{l\,m}\,{Y_{-1}}^{(l,m)}.
\end{eqnarray}
Thus the angular dependence of the dynamical equations is encoded in the spin weighted spherical harmonics. Furthermore, all the terms appearing in Eq.~(\ref{eq:d2Psi4}) have spin weight $-2$, and those appearing in Eq.~(\ref{eq:d2chi}) have spin weight $-1$, which confirms the equations are balanced with respect the spin weight.

The angular part of the perturbation equations can be factorized 
using the orthogonality properties of the spherical harmonics, and each equation and can be reduced to a set of equations for each mode as follows;
multiply Eq.~(\ref{eq:d2Psi4}) by ${Y_{-2}}^{(l',m')}$, and equation~(\ref{eq:d2Psi4}) by ${Y_{-1}}^{(l',m')}$. Then integrate the element of solid angle. 
Due to the 
orthogonality of the spherical harmonics
each sum
reduces and one gets an equation for each mode
($l, m$). 

Finally, recalling the {\it Peeling theorem} \cite{Wald84} which states that the Weyl scalars have the asymptotic decay $\Psi_s \equiv 1/r^{5-s}$, it proves  convenient to rewrite the equation for the gravitational perturbation in terms of the quantity $r\,{\Psi_4}^{(1)}$ which does not decay in the asymptotic region.
For the electromagnetic/gravitational perturbation the product between the Weyl and Maxwell scalars forming $\chi$ decays as $1/{r^4}$, thus the product $r^4 \chi$ has a constant amplitude.
For simplicity, let us define for each radial mode the amplitudes
\begin{equation}
{R_{l\,m}}_4=r\,{\Psi_{l\,m}}_4, \qquad {\rm and} \qquad {R_{l\,m}}_3=r^4\,\chi_{l\,m}. \label{eq:R4-R3}
\end{equation}
Then, the perturbation equations take the form 
%
\begin{eqnarray}
C_{1}\,{R_{l\,m}}_4 + C_{2}\,{R_{l\,m}}_3 &=&\kappa\,C_{3}\,\rho_{l\,m}\,
\left(k^\mu\,u_\mu\right)^2, \label{eq:d2Psi4a}\\
D_{1}\,{R_{l\,m}}_4 + D_{2}\,{R_{l\,m}}_3 &=&\kappa\,D_{3}\,\rho_{l\,m}\,
\left(k^\mu\,u_\mu\right)^2 - \frac{1}{4}\left(2\,\Phi_{11} + 3\,\Psi_{2}\right)\,
\frac{\sqrt{l\,\left(l+1\right)}}{\sqrt{2}\,r}\,{\rho_e}_{l\,m}\,\left(k^\mu\,u_\mu\right),
\label{eq:d2chia}
\end{eqnarray}
where
\begin{eqnarray}
C_1&=&\left[\left(\Delta - 3\,\mu\,\frac{6\,\Phi_{11} - 5\,\Psi_{2}}{2\,\Phi_{11} - 3\,\Psi_{2}}\right)\,\left(D + \rho - 4\,\epsilon\right) - 2\,\Phi_{11} + 3\,\Psi_2 - \frac{2\,\Phi_{11} - 3\Psi_{2}}{2\,\Phi_{11} + 3\Psi_{2}}\,\frac{\left(l-1\right)\,\left(l+2\right)}{2\,r^2} \right]\frac1r, 
 \nonumber \\
C_{2}&=&\sqrt{2\,\left(l-1\right)\,\left(l+2\right)}\,
\frac{\varphi_1}{2\,\Phi_{11} + 3\,\Psi_2}\,
\left(\left(\Delta - 3\,\mu\,\frac{6\,\Phi_{11} - 5\,\Psi_{2}}{2\,\Phi_{11} - 
3\,\Psi_{2}}\right)\,\frac{1}{r}\right)\,\frac{1}{r^4}, \nonumber \\
C_3&=-&\frac{2\,\Phi_{11} - 3\,\Psi_{2}}{2\,\Phi_{11} + 3\,\Psi_{2}}\frac{\sqrt{\left(l-1\right)\,l\,\left(l+1\right)\,\left(l+2\right)}}{2\,r^2},  \nonumber \\
D_1&=&\varphi_1\,\left[\sqrt{2\,\left(l-1\right)\,\left(l+2\right)}\,\frac{2\,\Phi_{11}}{2\,\Phi_{11} - 3\,\Psi_2}\,\left(D +  
\rho_s \, \frac{2\,\Phi_{11} + 9\,\Psi_2}{2\,\Phi_{11} + 3\,\Psi_2} - 4\,\epsilon\right)\,\frac{1}{r}\right]\,\frac{1}{r}, \nonumber \\
D_2&=&-\frac12\left[\left(D + 4\,\rho\,
\frac{\Phi_{11} + 3\,\Psi_2}{2\,\Phi_{11} + 3\,\Psi_2} - 4\,\epsilon\right)\,
\left(\Delta - 6\,\mu\right) + 2\,\left(2\,\Phi_{11} + 3\,\Psi_{2}\right) - 
\frac{\left(l-1\right)\,\left(l+2\right)}{2\,r^2}\frac{2\,\Phi_{11} + 3\,\Psi_{2}}{2\,\Phi_{11} - 3\,\Psi_{2}}\right]\,
\frac{1}{r^4}, \nonumber \\
D_3&=&\sqrt{\frac{l\,\left(l+1\right)}{2}}\,\varphi_1\,\left(D + 6\,\rho\,\frac{\Psi_{2}}{2\,\Phi_{11} + 3\,\Psi_2} - 4\,\epsilon\right)\,
\frac{1}{r}. \label{ops:CD}
\end{eqnarray} 
Eqs. (\ref{eq:d2Psi4a}, \ref{eq:d2chia}) are the final coupled dynamical equations for the 
gravitational and  electromagnetic perturbations in the time domain which can be written in any coordinate basis.

\section{Perturbations in Reissner-Nordstr\"om Space Time in Kerr Schild coordinates}
\label{sec:KS}

Following our previous work \cite{Moreno:2016urq}, we focus on the Reissner-Nordstr\"om metric in Kerr Schild-type coordinates:
\begin{equation}
ds^{2}=-\left(1-\frac{2\,M}{r}+\frac{Q^2}{r^2}\right)d{t}^{2}
+ 2\left(\frac{2\,M}{r}- \frac{Q^2}{r^2}\right)\,dt\,dr+\left(1+\frac{2\,M}{r}-\frac{Q^2}{r^2}\right)dr^{2}+
r^{2}\left( d\theta ^{2}+\sin ^{2}\theta d\varphi^{2}\right),  \label{eq:metric}
\end{equation}
where the null tetrad is
\begin{align}
l^{\mu}& =\frac{1}{2}\left( 1+\frac{2\,M}{r}-\frac{Q^2}{r^2}, 1-\frac{2\,M}{r}+\frac{Q^2}{r^2}, 0, 0\right),  \notag \\
k^{\mu}& = \left( 1,  -1, 0, 0\right),  \notag \\
m^{\mu}& =\frac{1}{\sqrt{2}\,r}\left( 0,0, 1, i\,\csc{\theta}  \right),
\end{align}
and the non-vanishing components of the Weyl, Ricci and Maxwell scalars  associated to this geometry are
\begin{equation}
\Psi _{2}=\frac{M}{r^3} -\frac{Q^2}{r^4}, \quad \Phi _{11}=\frac{Q^2}{2\,r^4}, \quad  \varphi_1=\frac{Q}{\sqrt{2}\,r^2}.
\label{weyl}
\end{equation}
For this spacetime and with our choice of null tetrad, the non zero spin coefficients are 
\begin{equation}
\mu  =\frac{1}{r}, \quad 
\rho_s =\frac{r^2-2Mr+Q^2}{2 r^3}, \quad
\epsilon = -\frac{1}{2}\left(\frac{M}{r^2}-\frac{Q^2}{r^3}\right), \quad 
\beta  =-\alpha =-\frac{1}{2\sqrt{2}}\frac{\cot \theta }{r}. \quad
\label{spinc}
\end{equation}

In order to solve the perturbation equations, the first step
is to write 
the equations in a dimensionless form.
For this purpose we recover the physical constants $G$, $c$, $\epsilon_0$ and $\mu_0$.
We start by defining the 
characteristic length of
the system $R_0$,
and write the radial coordinate $r$, as $r=R_0\,\hat{r}$, with $\hat{r}$ a dimensionless quantity. 
We also define a characteristic time, $T_0$, 
such that
$t=T_0\,\hat{t}$. In terms of the dimensionless quantities we find useful to define the quantity
\begin{equation}
 \sigma^2 = \frac{G\,M}{c^2\,R_0}.
 \label{eq:sigma2}
\end{equation}
The maximum value of the charge to get an extreme black hole is 
$Q_{\rm max}=\sqrt{\frac{G}{\mu_0}}\,\frac{M}{c}$, therefore we can normalize the value of the charge defining a dimensionless quantity   
$\hat{Q}=Q/Q_{\rm max}$.

The dimensionless electromagnetic and gravitational scalars are
\begin{eqnarray}
{\varphi_i}_g&=&\sqrt{\frac{G}{\mu_0}}\,\frac{1}{c^2}\,\varphi_i, \qquad \qquad \varphi_1=\frac{\sigma^2}{4\,\pi\,R_0}\,\frac{\hat{Q}}{\sqrt{2}\hat{r}^2}, \qquad \qquad
\chi=\frac{\sigma^2}{{R_0}^3}\,{\hat \chi} ,
\label{eq:phi_g}
\end{eqnarray}
where
\begin{equation}
 \hat \chi=3\,\left(\frac{1}{\hat r^3}
- \frac{\sigma^2 \, \hat Q^2}{\hat r^4}\right) \,\varphi_2^{(1)} - \frac{\hat Q^2}{2\,\sqrt{2}\,\pi\,\hat r^2}\Psi_3^{(1)},
\end{equation}
and:
\begin{eqnarray}
\Psi_2&=& \frac{\sigma^2}{{R_0}^2\, \hat{r}^4}\, \left(\hat{r} - \sigma^2\,\hat{Q}^2\right), \qquad \qquad
\Phi_{11}= \frac{\sigma^4}{{R_0}^2}\,\frac{\hat{Q}^2}{2\,\hat{r}^4}. \nonumber 
\label{eq:Psi2_adim}
\end{eqnarray}
Furthermore,
the dimensionless radial functions $\hat{R}_4,$ and $\hat{R}_3$ are
\begin{eqnarray}
\hat{R}_4=
\frac{1}{R_0} R_4\, , \qquad
\hat{R}_3= 
\frac{1}{R_0} R_3 \, .
\end{eqnarray}
Replacing the Weyl scalars Eq. \eqref{weyl}, and the spin coefficients Eq. (\ref{spinc}), in the perturbation equations Eqs. (\ref{eq:d2Psi4a}-\ref{eq:d2chia}).
We obtain the following system 
in terms of dimensionless quantities
\small{
\begin{eqnarray}
&&\left\{\left(\hat{r}^2 + 2\,\sigma^2\,\hat{r} - \sigma^4\,
\hat{Q}^2\right)\,\frac{\partial^2}{\partial\,\hat{t}^2} -
\left( \hat{r}^2 - 2\,\sigma^2\,\hat{r} + \sigma^4\,
\hat{Q}^2\right)\,\frac{\partial^2}{\partial\,\hat{r}^2} -
2\,\sigma^2\,\left(2\,\hat{r} - \sigma^2\,\hat{Q}^2\right)\,
\frac{\partial^2}{\partial\,\hat{t}\,\partial\,\hat{r}} + \right. \nonumber \\
&&  \left. - 2\,
\frac{\left(6\,\hat{r}^3 + (3 - 10\,\hat{Q}^2)\,\sigma^2\,\hat{r}^2  
- 5\,\sigma^4\,\hat{Q}^2\,\hat{r} - 
2\,\sigma^6\,\hat{Q}^4\right)}{\hat{r}\,(3\,\hat{r} - 4\,\sigma^2\,\hat{Q}^2)}\,\frac{\partial}{\partial \hat{t}} 
- 2\,
\frac{\left(6\,\hat{r}^3 - (3 - 10\,\hat{Q}^2)\,\sigma^2\,\hat{r}^2  
+ 5\,\sigma^4\,\hat{Q}^2\,\hat{r} + 
2\,\sigma^6\,\hat{Q}^4\right)}{\hat{r}\,(3\,\hat{r} - 4\,\sigma^2\,\hat{Q}^2)}\,\frac{\partial}{\partial \hat{r}}  \right. 
\nonumber \\
&& + \left. \left(l-1\right)\,\left(l+2\right)\,\frac{3\,\hat{r} - 
4\,\sigma^2\,\hat{Q}^2}{3\,\hat{r} - 2\,\sigma^2\,\hat{Q}^2} 
- \frac{6\,\sigma^2\,\left(\hat{r} - 2\,\sigma^2\,\hat{Q}^2 \right)}{\hat{r}\,\left(3\,\hat{r} - 4\,\sigma^2\,\hat{Q}^2 \right)}\right\}\,\hat{R}_4(\hat{t},\hat{r})  
\label{ec:S1_ff} + \\  &&
\frac{\sqrt{\left(l-1\right)\,\left(l+2\right)}\,\hat{Q}}{2\,\pi\,
\left(3\,\hat{r} - 2\,\sigma^2\,\hat{Q}^2 \right)}\,
\left(\frac{\partial}{\partial \hat{t}} - \frac{\partial}{\partial \hat{r}} + 
\frac{4\,\sigma^2\,\hat{Q}^2}{\hat{r}\,\left(3\,\hat{r} - 4\,\sigma^2\,\hat{Q}^2 \right)}\right)\,\hat{R}_3(\hat{t},\hat{r})
= 
\nonumber \\ && 
3\,\sigma^2\,\hat{r}\,\sqrt{\left(l-1\right)\,l\,\left(l+1\right)\left(l+2\right)}\,
\frac{3\,\hat{r} - 4\,\sigma^2\hat{Q}^2}{3\,\hat{r} - 2\,\sigma^2\,\hat{Q}^2}\,\hat{\rho}\left(\hat{t},\hat{r}\right)\,\left(\hat{u}_\mu\,k^\mu\right)^2,
 \nonumber
\end{eqnarray}
and
\begin{eqnarray}
&&\frac{\sigma^4\,\hat{Q}^3\,\sqrt{\left(l-1\right)\,\left(l+2\right)}}
{2\,\pi\,\left(3\,\hat{r} - 4\,\sigma^2\,\hat{Q}^2\right)}\,
\left[\left(\hat{r}^2 + 2\,\sigma^2\,\hat{r} - \sigma^4\,
\hat{Q}^2\right)\,\frac{\partial}{\partial\,\hat{t}} + \left(\hat{r}^2 - 2\,\sigma^2\,\hat{r} + \sigma^4\,\hat{Q}^2 \right)\,\frac{\partial}{\partial\,\hat{r}} + \right. \nonumber \\
&& \left.
\frac{3\,\hat{r}^2\,\left(\hat{r} + 2\,\sigma^2\right)- 
\sigma^2\,\hat{Q}^2 \left(4\,\hat{r}^2 + 9\,\sigma^2\,\hat{r} -
\sigma^4\,\hat{Q}^2\right)}{\hat{r}\,\left(3\,\hat{r} - 2\,\sigma^2\,\hat{Q}^2 \right)}\right]\,\hat{R}_4(\hat{t},\hat{r}) \nonumber \\
&& + 
\left\{\left(\hat{r}^2 + 2\,\sigma^2\,\hat{r} - \sigma^4\,
\hat{Q}^2\right)\,\frac{\partial^2}{\partial\,\hat{t}^2} -
\left( \hat{r}^2 - 2\,\sigma^2\,\hat{r} + \sigma^4\,
\hat{Q}^2\right)\,\frac{\partial^2}{\partial\,\hat{r}^2} -
2\,\sigma^2\,\left(2\,\hat{r} - \sigma^2\,\hat{Q}^2\right)\,
\frac{\partial^2}{\partial\,\hat{t}\,\partial\,\hat{r}} \right. \nonumber \\
&&  \left. - 2\,\frac{3\,\hat{r}^3 - \sigma^2\,\hat{Q}^2\,\hat{r}^2 + 
\sigma^4\,\hat{Q}^2\,\hat{r} - \sigma^6\,\hat{Q}^4}{\hat{r}\,\left(3\,\hat{r} - 2\,\sigma^2\hat{Q}^2 \right)}\,
\frac{\partial}{\partial\,\hat{t}}
- 2\,\frac{ \left(\hat{r} - \sigma^2\,\hat{Q}^2 \right) \,
\left(3\,\hat{r}^2 - \sigma^4\,\hat{Q}^2 \right)}{\hat{r}\,\left(3\,\hat{r} - 2\,\sigma^2\hat{Q}^2 \right)}\,\frac{\partial}{\partial\,\hat{r}}  
 \right. \nonumber \\ 
 && 
\left. + \left(l-1\right)\,\left(l+2\right)\,\frac{3\,\hat{r} - 
2\,\sigma^2\,\hat{Q}^2}{3\,\hat{r} - 4\,\sigma^2\,\hat{Q}^2} 
+ 2\,\frac{3\,\hat{r}^2 - \sigma^4\,\hat{Q}^2}{\hat{r}\,\left(3\,\hat{r} - 
2\,\sigma^2\,\hat{Q}^2\right)}\right\}  
\,\hat{R}_3(\hat{t},\hat{r}) = \label{ec:S2_ff} \\ 
&&- \frac{3\,\sigma^4\,\hat{r}\,\sqrt{l\,\left(l+1\right)}\hat{Q}}{2\,\pi}\,\left[\left(r^2 + 2\,M \, r - Q^2 \right)\,
\frac{\partial}{\partial\,\hat{t}} +  
\left(r^2 - 2\,M \, r + Q^2 \right)\,\frac{\partial}
{\partial\,\hat{r}}  
\right. \nonumber \\
&& 
\left. + 
\frac{2\,\hat{r}^3 + \sigma^2\,\left(8-\hat{Q}^2\right)\,\hat{r}^2 - 16\,\sigma^4\,\hat{Q}^2\hat{r} + 7\,\sigma^6\,\hat{Q}^4}{\hat{r}\,\left(3\,\hat{r}
- 2\,\sigma^2\,\hat{Q}^2\right)}
\right]\,\hat{\rho}(\hat{t},\hat{r})\,\left(\hat{u}_\mu\,k^\mu\right)^2  
- \frac{3\,\hat{r}\,\sqrt{2\,l\,\left(l+1\right)}}
{4\,\pi}\,\left(3\,\hat{r} - 2\,\sigma^2\,\hat{Q}^2\right)\,
\hat{\rho}_e\,{{\hat{u_e}}}_\mu\,k^\mu.
\nonumber
\end{eqnarray}}
where for simplicity in the notation, we have drop the indices $l,m$ on each mode.

\subsection{Matter models} \label{sec:Sources}
We shall consider that there are two sources of matter that cause a perturbation into the black hole. One associated to neutral
matter characterized by the rest mass density $\hat{\rho}_n$, and the other associated to charged particles with density $\hat{\rho}_e$. 
We consider that neutral particles move only in the radial direction and are free falling into the black hole.
The conservation of the number of particles $\nabla_\mu J^\mu=0$, 
where $J^\mu=\rho_n u^\mu_n$, for the metric Eq. \eqref{eq:metric} gives:
\begin{equation}
\frac{\partial\,\hat{\rho}_n}{\partial\,\hat{t}} + v_n^r\,
\frac{\partial\,\hat{\rho}_n}{\partial\,\hat{r}} + 
\hat{\rho}_n\,\frac{4\,(\hat{u}_n^r)^2 + \hat{r}\,
\frac{\partial\,(\hat{u}_n^r)^2}{\partial\,\hat{r}}}{2\,\hat{r}\,(\hat{u}_n^r)^2}
=0, \label{eq:cont}
\end{equation}
where we have defined the 3-velocity $v_n^r = \frac{\hat{u}_n^r}{\hat{u}_n^0}$. 
and we have assumed that the dimensionless four velocity has components ${\hat u}_n^\mu=(\hat{u}_n^0, \hat{u}_n^r, 0,0)$.
For charged particles we make the same assumptions and obtain that the conservation of the number of particles implies
\begin{equation}
\frac{\partial\,\hat{\rho}_e}{\partial\,\hat{t}} + v_e^r\,
\frac{\partial\,\hat{\rho}_e}{\partial\,\hat{r}} + 
\hat{\rho}_e\,\frac{4\,(\hat{u}_e^r)^2 + \hat{r}\,
\frac{\partial\,(\hat{u}_e^r)^2}{\partial\,\hat{r}}}{2\,\hat{r}\,(\hat{u}_e^r)^2}
=0. \label{eq:cont2}
\end{equation}

Considering the normalization of the four velocity 
$\hat u^\mu\,\hat u_\mu=-1$ for both types of particles,
we
can express $\hat{u}^0$ in terms of $\hat{u}^r$ as \cite{Moreno:2016urq}:
\begin{equation}
\hat{u}^0=\frac{\sigma^2\,\left(2 - \frac{\sigma^2\,{\hat{Q}}^2}{\hat{r}}\right)\,\frac{\hat{u}^r}{\hat{r}} + 
\frac{1}{\sigma}\,\sqrt{\sigma^2\,(\hat{u}^r)^2 - k\,\hat{\Delta}}}{\hat{\Delta}}, \label{eq_u0_g} 
\end{equation}
where $\hat{\Delta}=1 - \frac{2\,\sigma^2}{\hat{r}} + \frac{\sigma^4\,\hat{Q}^2}{\hat{r}^2}$. As previously stated, this expression is valid for neutral and charged particles.

For spherically symmetric static spacetimes one can determine the movement of test particles up to quadrature by means of the Euler Lagrange equations as follows.
Let us consider the dimensionless lagrangian for neutral particles
\begin{eqnarray}
\hat{\mathcal{L}}&=\frac12\,&\left(g_{\mu\nu}\,
 \hat{u}^{\mu}\,\hat{u}^{\nu} \right), \label{eq:lagrangian_n}
 \end{eqnarray}
using the Euler Lagrange equation for $x^{0}$, and the metric Eq. (\ref{eq:metric}),
one obtains a conserved quantity associated to the energy of the particles at infinity $\hat{\epsilon}_n$ and consequently 
\begin{eqnarray}\label{eq_u0n}
{\hat u}_n^0&=&\frac{\hat{\epsilon}_n + \sigma^2\,
\left(2 - \frac{\sigma^2\,\hat{Q}^2}{\hat{r}}\right)\,\frac{\hat{u}^r}{\hat{r}}}{\hat{\Delta}},
\end{eqnarray}
where $\hat{\epsilon}_n=-\frac{\partial\,\hat{\mathcal{L}}}{\partial\,\hat{u}^0}$.
Proceeding in a similar way for charged particles, the dimensionless lagrangian is 

%
%
\begin{equation}
 \hat{\mathcal{L}}_e=\frac{1}{2}
\left(\hat{u}^\mu\,\hat{u}_\mu + 
2\,\sigma^2\,\hat{q}\,\hat{Q}\,\frac{M}{m}\,\frac{\hat{u}^0}{\hat{r}}\right),
\end{equation}
where the vector potential was taken as $A_\mu=\left(\frac{Q}{\epsilon_0\,r},\vec{0}\right)=
\left(\frac{Q_{\rm max}\,\mu_0\,c^2}{R_0}\,\frac{\hat{Q}}{\hat{r}},\vec{0}\right)$ and we have written $q=Q_{\rm max}\,\hat{q}$. 
From the staticity of the space time via the Euler lagrange equation for $t$ we get
\begin{eqnarray}\label{eq_u0e}
{\hat u}_e^0&=&\frac{\hat{\epsilon}_e + \sigma^2\,
\left(2 - \frac{\sigma^2\,\hat{Q}^2}{\hat{r}}\right)\,\frac{\hat{u}^r}{\hat{r}} - \frac{\lambda}{2\,\hat{r}}}{\hat{\Delta}},
\end{eqnarray}
where $\hat{\epsilon}_e=-\frac{\partial\,\hat{\mathcal{L}}_e}{\partial\,\hat{u}^0}$ and $\lambda=2\,\sigma^2\,\hat{q}\,\hat{Q}\,\frac{M}{m}$.
By using Eq. \eqref{eq_u0_g} in both Eqs. (\ref{eq_u0n}, \ref{eq_u0e}) we obtain

\begin{eqnarray}
\left(\hat{u}^r_e\right)^2&=&\left(\hat{\epsilon}_e - \frac{\lambda}{2\,\hat{r}}\right)^2 - \frac{\hat{\Delta}}{\sigma^2}, \\
\left(\hat{u}_n^r\right)^2&=&{\hat{\epsilon}_n}^2 - \frac{\hat{\Delta}}{\sigma^2}.
\end{eqnarray}

Finally, the projection of the four velocity along the null vector $k^\mu\,u_\mu$ for both types of particles
provides a couple of expressions that  will be useful bellow:
${\hat{u}^0}_e + {\hat{u}^r}_e=\frac{\hat{\epsilon}_e - \frac{\lambda}{2\,\hat{r}} + \hat{u}^r}{\hat{\Delta}}$, and ${\hat{u}^0}_n + {\hat{u}^r}_n=\frac{\hat{\epsilon}_n + \hat{u}^r}{\hat{\Delta}}$.

Given the components of the velocity,
for both neutral and charged particles and 
the evolution for the densities one can solve 
the perturbation equations numerically given an initial distribution of matter and compute the resultant gravitational signal due to the gravitational and electromagnetic perturbation induced into the black hole.

\subsection{First order reduction and numerical implementation}

In terms of the 3+1 decomposition of  the  spacetime 
$ds^2=-(\alpha^2 -\beta^r\beta_r)dt^2+2\,\beta_r dr dt+\gamma_{rr}^2 dr^2 + r^2(d\theta^2+\sin^2 d \phi^2)$ one can determine the lapse function $\alpha$, the shift vector $\beta^r$ and the metric component $\gamma_{rr}$,
from the metric Eq.~(\ref{eq:metric}) which gives
\begin{eqnarray}
 \alpha=\left(1+\frac{2\,\sigma^2}{\hat{r}}-\frac{\sigma^4\,\hat{Q}^2}{\hat{r}^2} \right)^{-1/2} \, , 
 \qquad \beta_{r} =\frac{2\,\sigma^2}{\hat{r}}-\frac{\sigma^4\,\hat{Q}^2}{\hat{r}^2}\, ,
 \qquad \beta^{r}=\alpha^2\beta_{r} \, , 
 \qquad \gamma_{rr}=\left(1-\frac{2\,\sigma^2}{\hat{r}}+\frac{\sigma^4\,\hat{Q}^2}{\hat{r}^2} \right)^{1/2}.
\end{eqnarray}
In order to write the system of second order differential equations Eqs.~(\ref{eq:d2Psi4a}-\ref{eq:d2chia}) as a first order system suitable for numerical integration, we introduce the 
auxiliary functions
%
\begin{equation}
\hat{\pi}_a \equiv \frac{1}{\alpha^2}\,
\left( \partial_{\hat{t}}\,\hat{R}_a - \beta^r\,\hat{\psi}_a \right), 
\qquad 
\psi_a \equiv \partial_{\hat{r}}\,\hat{R}_a, \qquad {\rm with} \quad a=3,4.
\end{equation}
From the definition of $\hat{\pi}_a $ one obtains the time development of $\hat R_a$, 
while by taking the time derivative of $\hat \psi_a$ and replacing the lapse and shift one gets %
\begin{eqnarray}\label{fo1}
\partial_{\hat{t}}\hat{R}_a &=& \frac{\hat{r}^2\hat{\pi}_a + 
\sigma^2\left(2\hat{r} - \sigma^2\hat{Q}^2\right)\hat{\psi}_a}{\hat{r}^2 + 2\sigma^2\hat{r} - \sigma^4\,\hat{Q}^2}, \\
 \partial_{\hat{t}}\hat{\psi}_a &=& \frac{\hat{r}^2\partial_{\hat{r}}
 \hat{\pi}_a + \sigma^2\left(2\,\hat{r} - \sigma^2\,\hat{Q}^2\right)\,\partial_{\hat{r}}\hat{\psi}_a}{\hat{r}^2 + 2\sigma^2\hat{r} -
 \sigma^4\hat{Q}^2}  + \frac{2\hat{r}\sigma^2\,\left(
 \hat{r} - \sigma^2\hat{Q}^2\right)}
 {\left(\hat{r}^2 + 2\sigma^2\hat{r} - 
 \sigma^4\hat{Q}^2\right)^2}\left(\hat{\pi}_a - \hat{\psi}_a\right).
\end{eqnarray} 
The equations for $\hat \pi_4$ and $\hat \pi_3$ are obtained by substituting first order functions into Eqs. (\ref{ec:S1_ff}-\ref{ec:S2_ff}),
\begin{eqnarray}
&&\partial_{\hat{t}}\,\hat{\pi}_4 = \frac{\sigma^2\,\left(2\,\hat{r} -
\sigma^2\,\hat{Q}^2\right) \partial_{\hat{r}}\,\hat{\pi}_4  + 
\hat{r}^2\,\partial_{\hat{r}}\,\hat{\psi}_4 }{\left(
\hat{r}^2 + 2\,\sigma^2\,\hat{r} - \sigma^4\,\hat{Q}^2 \right)}  \nonumber \\
&& + 2\frac{3\,\hat{r}^3\,\left(2\,\hat{r}^2+5\,\sigma^2\,\hat{r} + 4\,\sigma^4 \right)- 
\sigma^2\,Q^2\,\left(10\,\hat{r}^4 +31\,\sigma^2\,\hat{r}^3 + 2\,\sigma^4\,\left(15 - 4\,\hat{Q}^2\right)\,\hat{r}^2 - 2\,\sigma^6\,\hat{Q}^2\,\left(8\,\hat{r}-\sigma^2\,\hat{Q}^2\right) \right)}{\hat{r}\,\left(\hat{r}^2 + 2\,\sigma^2\,\hat{r} - \sigma^4\,\hat{Q}^2 \right)^2\, \left(3\,\hat{r} - 4\,\sigma^2\,\hat{Q}^2 \right)}\,\hat{\pi}_4 \nonumber \\
&&+ 2\frac{3\,\hat{r}^3\,\left(2\,\hat{r}^2+11\,\sigma^2\,\hat{r} + 12\,\sigma^4 \right) + 
\sigma^2\,Q^2\,\left(10\,\hat{r}^4 +7\,\sigma^2\,\hat{r}^3 - 2\,\sigma^4\,\left(9 + 4\,\hat{Q}^2\right)\,\hat{r}^2 - 2\,\sigma^6\,\hat{Q}^2\,\left(2\,\hat{r}-\sigma^2\,\hat{Q}^2\right) \right)}{\hat{r}\,\left(\hat{r}^2 + 2\,\sigma^2\,\hat{r} - \sigma^4\,\hat{Q}^2 \right)^2\, \left(3\,\hat{r} - 4\,\sigma^2\,\hat{Q}^2 \right)}\,\hat{\psi}_4 \nonumber \\
&& -\left(\frac{(l-1)(l+2) \, \left(3\,\hat{r} - 4\,\sigma^2\,\hat{Q}^2 \right)}{\hat{r}^2\, \left(3\,\hat{r} - 2\,\sigma^2\hat{Q}^2 \right)} -\frac{6\,\sigma^2\,\left(\hat{r} - 
2\,\sigma^2\,\hat{Q}^2 \right)}{\hat{r}^3\,\left(3\,\hat{r} - 4\,\sigma^2\,\hat{Q}^2 \right)}\right)\,\hat{R}_4 \nonumber \\
&& - \frac{\hat{Q}\,\sqrt{(l-1)(l+2)}}{2\,\pi\,\left(3\,\hat{r} - 2\,\sigma^2\,\hat{Q}^2 \right)} \left(\frac{\hat{\pi}_3-\hat{\psi}_3}{\left(\hat{r}^2 + 2\,\sigma^2\,\hat{r} - 
\sigma^4\,\hat{Q}^2\right)} + 4\,\frac{\sigma^2\,\hat{Q}^2}{\hat{r}^3\,\left(3\,\hat{r} - 
4\,\sigma^2\,\hat{Q}^2 \right) }\,\hat{R}_3 \right) \nonumber \\
&& 
+3\,\sigma^2\,\sqrt{\left(l-1\right)\,l\,\left(l+1\right)\left(l+2\right)}\,
\frac{3\,\hat{r} - 4\,\sigma^2\,\hat{Q}^2 }{\hat{r}\,\left(3\,\hat{r} - 
2\,\sigma^2\,\hat{Q}^2\right)}\,\hat{\rho_n}(\hat{t},\hat{r})\,(\hat{u_n}^0+\hat{u_n}^r)^2. 
\label{R3}
\end{eqnarray}
\begin{eqnarray}
&&\partial_{\hat{t}}\,\hat{\pi}_3 = \frac{\sigma^2\,\left(2\,\hat{r} - \sigma^2\,\hat{Q}^2\right)\,\partial_{\hat{r}}\,\hat{\pi}_3  + \hat{r}^2\,\partial_{\hat{r}}\,\hat{\psi}_3}{\hat{r}^2 + 
2\,\sigma^2\,\hat{r} - \sigma^4\,\hat{Q}^2}  \nonumber \\
&& + 2\,\frac{3\,r^3\,\left(r^2+ 2\,\sigma^2\,\hat{r} + 2\,\sigma^4 \right) - 
\sigma^2\,\hat{Q}^2\,\left(\hat{r}^4 + 
4\,\sigma^2\,\hat{r}^3 + 11\,\sigma^4\,\hat{r}^2 - \sigma^6\,\hat{Q}^2\,\left(6\,\hat{r} - 
\sigma^2\,\hat{Q}^2\right) \right)}{\hat{r}\, \left(\hat{r}^2 + 2\,\sigma^2\,\hat{r} - 
\sigma^4\,\hat{Q}^2\right)^2\,\left(3\,\hat{r} - 2\,\sigma^2\,\hat{Q}^2\right)}\,\hat{\pi}_3 
\nonumber \\
&& + 2\,\frac{3\,r^3\,\left(r^2+ 6\,\sigma^2\,\hat{r} + 6\,\sigma^4 \right) - 
\sigma^2\,\hat{Q}^2\,\left(3\,\hat{r}^4 + 
24\,\sigma^2\,\hat{r}^3 + \left(29-8\,\hat{Q}^2\right)\,\sigma^4\,\hat{r}^2 - \sigma^6\,\hat{Q}^2\,\left(16\,\hat{r} - 3\,\sigma^2\,\hat{Q}^2\right) \right)}{\hat{r}\, \left(\hat{r}^2 + 2\,\sigma^2\,\hat{r} - 
\sigma^4\,\hat{Q}^2\right)^2\,\left(3\,\hat{r} - 2\,\sigma^2\,\hat{Q}^2\right)}\,\hat{\psi}_3 \nonumber \\
&& - \left(\frac{(l-1)(l+2) \, \left(3\,\hat{r} - 2\,\sigma^2\,\hat{Q}^2\right)}{r^2\, \left(3\,\hat{r} - 
4\,\sigma^2\,\hat{Q}^2\right)} + 
\frac{2\,\left(3\,\hat{r}^2 - \sigma^4\,\hat{Q}^2\right)}{\hat{r}^3\,\left(3\,\hat{r} - 
2\,\sigma^2\,\hat{Q}^2\right)}\right)\,\hat{R}_3 \label{R4} \\
&& - \frac{\sigma^4\,\hat{Q}^3\,\sqrt{(l-1)(l+2)}}{2\,\pi\,\left(3\,\hat{r} - 4\,\sigma^2\,\hat{Q}^2\right)}
\,\left(\hat{\pi}_4 + \hat{\psi}_4  + \frac{3\,\hat{r}\,\left(\hat{r}^2 + 2\,\sigma^2\,\hat{r} -
3\,\sigma^4\,\hat{Q}^2\right)- \sigma^2\,\hat{Q}^2\,\left(4\,\hat{r}^2 - \sigma^4\,\hat{Q}^2\right)}{\hat{r}^3\,
\left(3\,\hat{r} - 2\,\sigma^2\hat{Q}^2\right)}\,\hat{R}_4\right) \nonumber \\
&& + \frac{3\,\sigma^4\,\hat{Q}\,\sqrt{l\,\left(l+1\right)}}{2\,\pi\,\hat{r}}\,\left(\hat{u}^t+\hat{u}^r\right)^2\,\left[
\left(\left(\hat{r}^2 + 2\,\sigma^2\,\hat{r} - \sigma^4\,\hat{Q}^2\right)\,
\partial\,_{\hat{t}}\,\hat{\rho} + 
\left(\hat{r}^2 - 2\,\sigma^2\,\hat{r} + \sigma^4\,\hat{Q}^2\right)\,
\partial\,_{\hat{r}}\,\hat{\rho}\right)
\right. \nonumber \\
&& \left. + \hat{\rho}\,\left(2\,\left(\hat{r}^2 - 2\,\sigma^2\,\hat{r} + \sigma^4\,\hat{Q}^2\right)\,\frac{\partial\,_{\hat{r}}\,\left(\hat{u}^t+\hat{u}^r\right)}{\left(\hat{u}^t+\hat{u}^r\right)} + \frac{2\,\hat{r}^3 + \sigma^2\,\left(8-\hat{Q}^2\right)\,\hat{r}^2 - 
16\,\sigma^4\,\hat{Q}^2\,\hat{r} + 7\,\sigma^6\,\hat{Q}^4}{\hat{r}\,\left(3\,\hat{r}-2\sigma^2\,\hat{Q}^2\right)}\right)\right] \nonumber + \\
&&\frac{3\,\sigma^4\,\hat{Q}\,\sqrt{2\,l\,\left(l+1\right)}\,\left(3\,\hat{r}-2\sigma^2\,\hat{Q}^2\right)}{4\,\pi\,\hat{r}}\,\hat{\rho}_e\,\left(\hat{u_e}^0+\hat{u_e}^r\right). \nonumber
\end{eqnarray}
%
\subsection{Waveforms}

For our simulations we solve the system of equations for the gravitational perturbation and the electromagnetic/gravitational perturbation with sources Eqs. (\ref{R3}-\ref{R4}). For the source we are considering non spherical shell of particles falling into the hole. The numerical code evolves the first order variables with a third order Runge Kutta integrator with a fourth order spatial stencil. For more detailed description of the code see Ref. \cite{Moreno:2016urq}.
We use as initial data a gaussian packet in the density describing a non spherical shell of particles falling into the hole $\rho_n(0,\hat{r})=\rho_0 e^{(\hat{r}-\hat{r}_{cg})^2/2\sigma_g^2}=\rho_e(0,\hat{r})$; where $\rho_0=5 \times 10^{-3}$, $\hat{r}_{cg}=50$ and $\sigma_g=0.8$. For the simulation we are using Kerr–Schild type coordinates and $r_{min}$ lies inside the event horizon, we choose $\hat r_{\rm min}=1.5$ and $\hat r_{\rm max}=1000$. The gravitational and electromagnetic/gravitational waveforms produced by the infalling matter are extracted at a fixed $\hat{r}=100$ radius. The gravitational and electromagnetic/gravitational functions, $\hat R_4$, $\hat R_3$, initially are set to zero, as well as their time derivatives; the outer boundary was set far enough from the horizon to ensure that any possible incoming radiation has no effect on our results. 

\begin{figure}
    \begin{tabular}{cc}
    \includegraphics[width=0.49 \textwidth]{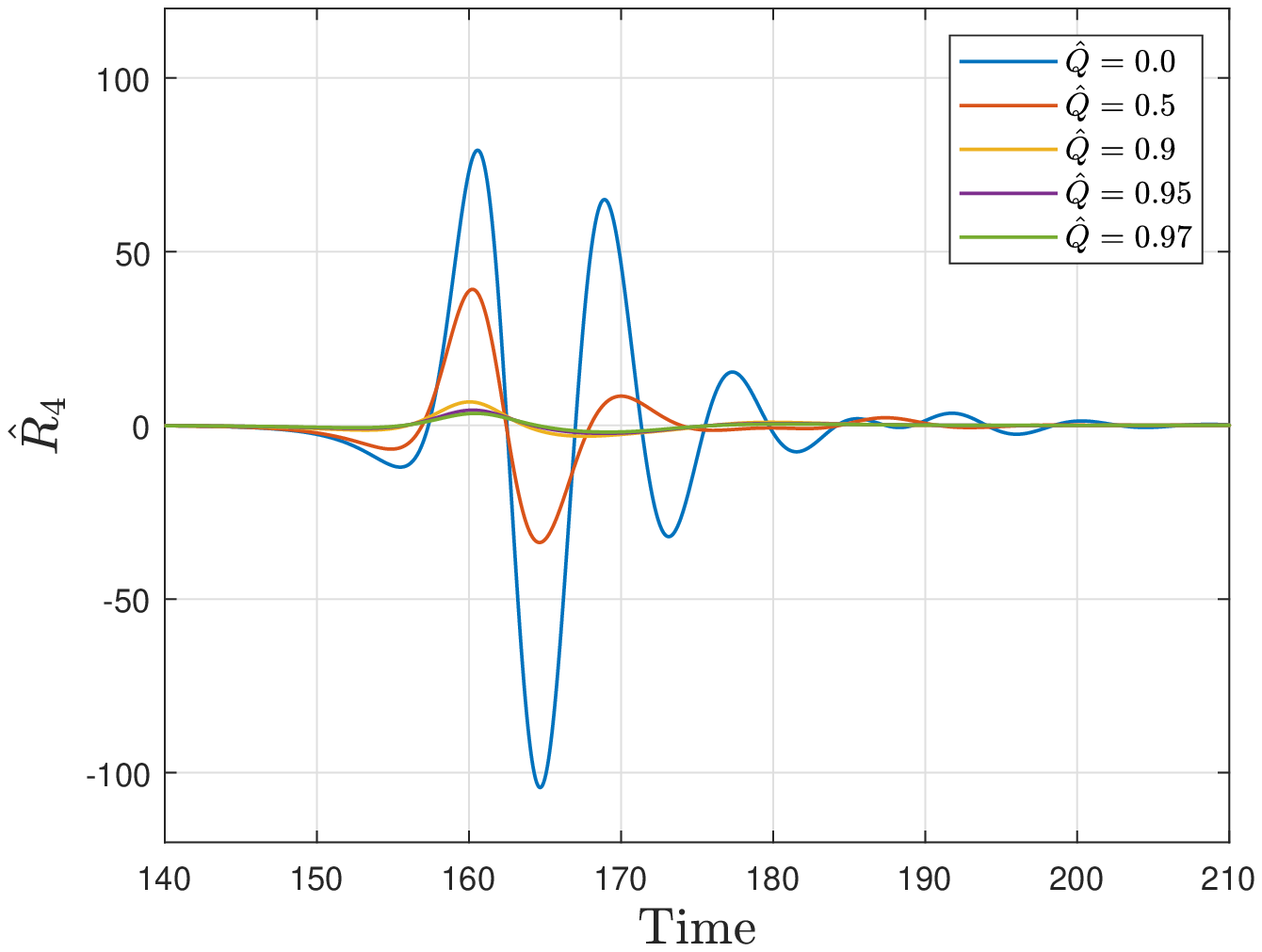}
    & 
    \includegraphics[width=0.49 \textwidth]{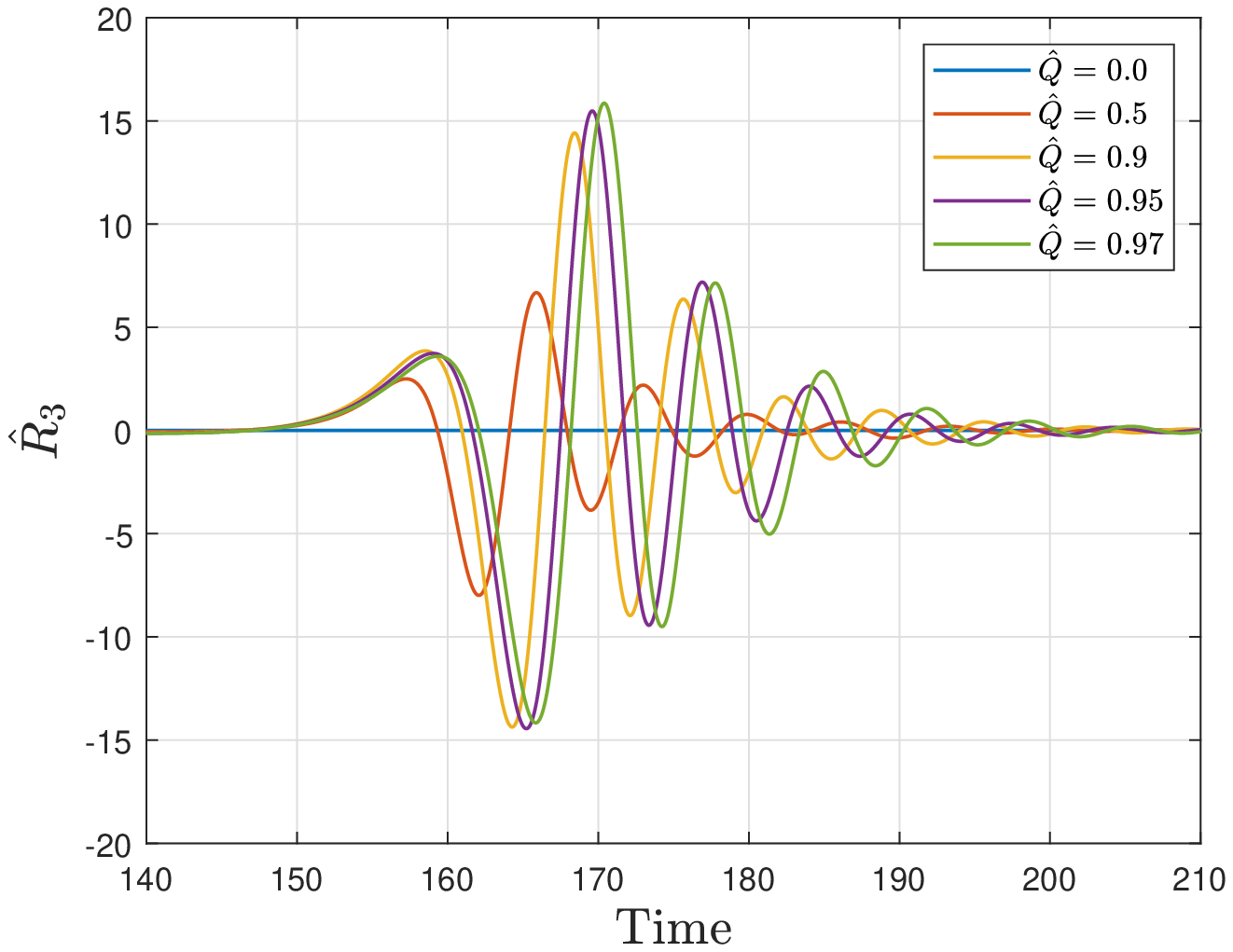}
    \\
    (a) & (b)
    \end{tabular}
    \caption{Radial profiles of the gravitational $\hat R_4$ (a), and electromagnetic/gravitational $\hat R_3$ (b) signals for the quadrupolar $l = 2$ mode with different values of $\hat Q = 0.0, 0.5, 0.9, 0.95$ and $0.97$.}
    \label{fig:R3vsR4}
\end{figure}

In Fig. \ref{fig:R3vsR4} we show the radial profile for the gravitational (a) and the electromagnetic/gravitational component (b) for the mode $l=2$. The simulation have been done for different values of charge $\hat Q = 0.0, 0.5, 0.9, 0.95$ and $0.97$. We can observe that the waveforms are quiet similar in structure and that the amplitude changes for different values of $\hat Q$; the gravitational waveform amplitude is greater and present more oscillations that the electromagnetic/gravitational case. It is interesting to observe that the amplitude decreases for greater values of $\hat Q$ in the gravitational case; on the other hand, in the electromagnetic/gravitational case the amplitude increase for greater values of $\hat Q$. The time response for both, gravitational and electromagnetic/gravitational waveforms stars and ends almost at similar times. 

\begin{figure}
    \begin{tabular}{cc}
    \includegraphics[width=0.49 \textwidth]{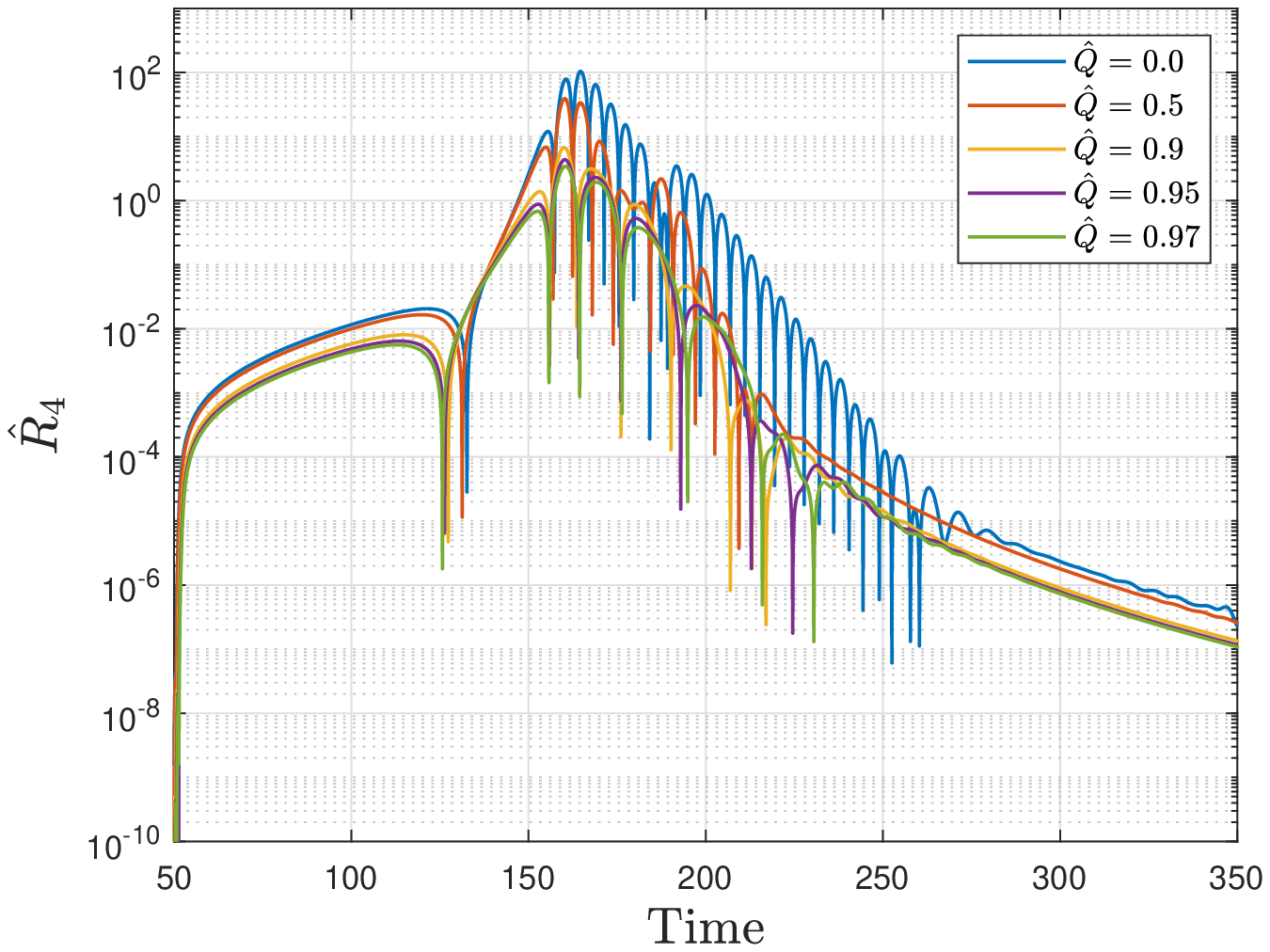}
    & 
    \includegraphics[width=0.49 \textwidth]{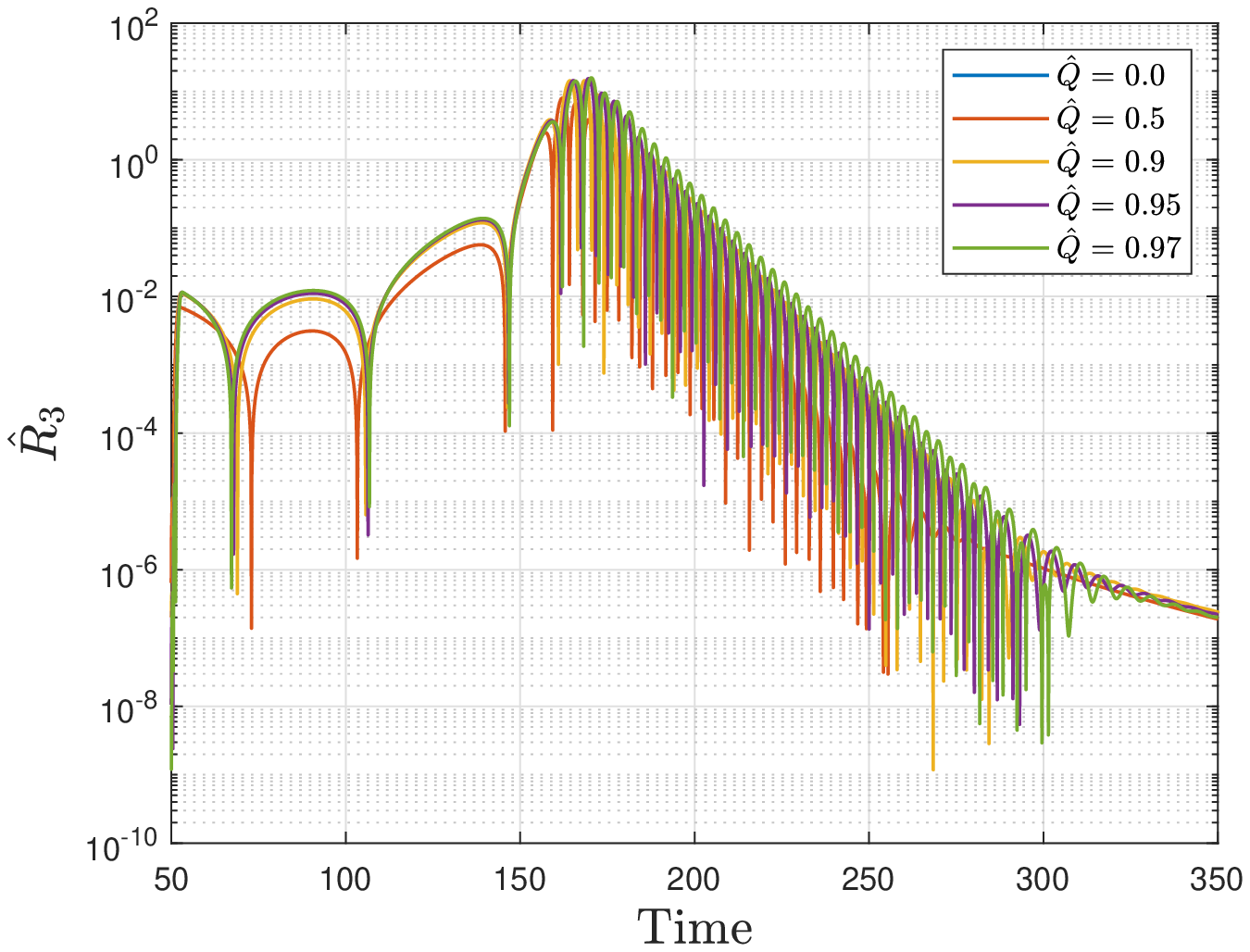}
    \\
    (a) & (b)
    \end{tabular}
    \caption{The gravitational and electromagnetic/gravitational of the signals $\hat R_4$ and $\hat R_3$ 
    in a logarithmic scale for different values of charge $\hat Q = 0.0, 0.5, 0.9, 0.95$ and $0.97$. The signals show the characteristics phases: initial burst, the quasinormal ringing, the power-law decay and the tail.}
    \label{fig:abs}
\end{figure}

In Fig. \ref{fig:abs} we plot the absolute value on a logarithmic scale to show the different stages of the signal: the initial burst due to the initial data, the quasinormal ringing and the tail. The left figure (a) for the gravitational signal and (b) for the electromagnetic/gravitational signal.  We find no sign of mixing between gravitational and electromagnetic/gravitational frequencies. Each signal displays its characteristic ring down frequency and the power-law decay. The ring down frequencies has been extracted from $\hat t \sim 50$ to $\hat t \sim 350$ for the gravitational case and from $\hat t \sim 50$ to $400$ for the electromagnetic/gravitational case.
The frequency of the gravitational waves is the one associated to the quadrupolar quasinormal mode. In order to find the frequencies we fit the data with a sinusoidal waveform. The numerical values of the corresponding frequencies are shown in table \ref{tab:frequency}. The values we find are consistent with the values given in \cite{Chandrasekhar83}. Although it is known that quasinormal mode frequencies are complex, we were interested on the oscillatory behaviour of the signal. For black holes of with masses between
$10M_{\odot}< M <10^3M_{\odot}$ the electromagnetic frequencies are in the interval $8$\,Hz to $800$\,Hz whereas the gravitational waves produced for such a range of masses are in the interval of $12$Hz to $1.2$kHz \cite{Moreno:2016urq}. As has been pointed out, quasinormal ringing can be used to determine the intrinsic properties of the black hole \cite{Kokkotas:2010zd}. Electromagnetic waves with such low frequencies however could be absorbed easily by the interstellar medium during its propagation and it will be almost impossible to detect them directly.

\begin{table}
		\begin{tabular}{|c|c|c|c}  
			\hline 
			$\hat{Q}$&$\ell$&$\ \hat{\omega}$ \\
			\hline
			$0$ & 2 & $0.3736$ \\
			$0.2$ & 2&$0.3747$ \\
			$0.9$ & 2&$0.4135$ \\
            $0.95$ & 2&$0.4216$ \\
			
			\hline 
        \end{tabular}
	\caption {Frequency of the quasinormal modes of the black hole produced by the perturbation of the accreting matter.The frequencies are consistent with the values given in \cite{Chandrasekhar83}}\label{tab:frequency}
\end{table}

Next we present the behavior of the energy carried by the gravitational wave for several values of the charge of the infalling matter for the $l=2$ mode (any $-l \leq m \leq l$), obtained by from the energy loss formula, that is, the power of the gravitational wave, $P_{\rm gw}=\frac{dE_{\rm gw}}{dt}$, (see \cite{Degollado:2009rw}):
\begin{equation}
P_{\rm gw}
=\lim_{r\to \infty}\frac{1}{16\pi}\,\int_{-\infty}^t|R_4|^2\,dt'.
\label{eq:dEdt_R4}
\end{equation}
%
%

\begin{figure}
    \includegraphics[width=0.49 \textwidth]{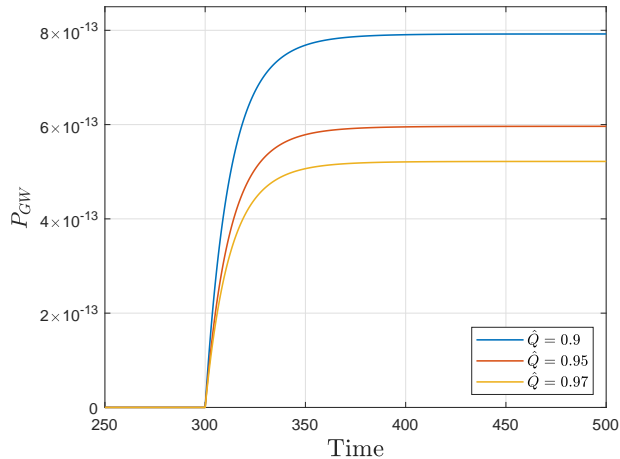}
    \caption{Energy carried by the gravitational wave, $\Psi_4^{(1)}$, according to Eq.~(\ref{eq:dEdt_R4}), for the values $\hat{Q}=0.9, 0.95, 0.97$ proportional to the charge of the black hole.}
    \label{fig:ener_R4}
\end{figure}

From Fig.~\ref{fig:ener_R4}, we notice how the flux of energy 
carried by the gravitational wave $\Psi_4^{(1)}$, reaches a constant value.
This value decreases
for large values of the charge of the black hole.
Although the same integral can be made for the gauge invariant quantity $\chi$, its interpretation as electromagnetic energy is not immediate, as long as it is coupled to the gravitational radiation through $\Psi_3^{(1)}$, and deserves a deeper discussion. Here we only mention that it is zero for $\hat{Q}=0$ and opposite in behaviour to $dE_{\rm gw}/dt$; its asymptotic value increases with $\hat{Q}$.

\begin{figure}
    \includegraphics[width=0.49 \textwidth]{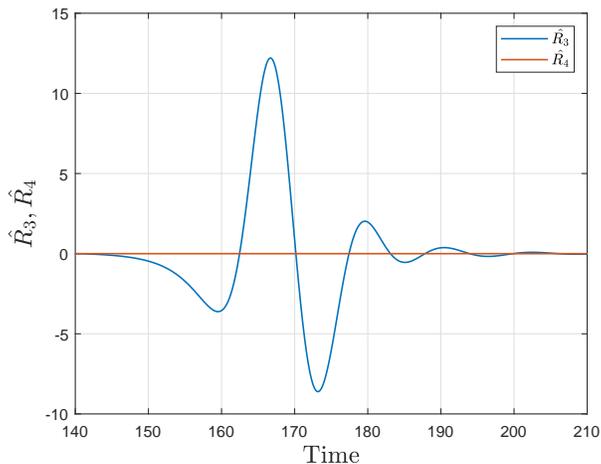}
    \caption{Wave forms for a dipolar perturbation, $l=1$, with $\hat{Q}=0.9$, 
    $M=1$ and $\hat{q}=-0.8$, data are collected at $r_{\rm obs}=100$. Notice how in this case, there is no gravitational response, $\Psi_4^{(1)}=0$, as it should be, but there do is the response associated with $\chi$.}
    \label{fig:l1_R3_R4}
\end{figure}

Our system of equations allows to have a purely electromagnetic/gravitational response (encoded in the function $\chi$). Indeed, for the $l=1$ case, which corresponds to a dipole angular distribution, there is no gravitational response, as the gravitational waves occur starting from the quadrupolar angular distribution \cite{Schutz80}, but there might be an electromagnetic one. We see that this is the case; solving the system of equations with $l=1$ we obtain only a response in $\chi$, which we present in Figure \ref{fig:l1_R3_R4}. 

%

\section{Final remarks}
\label{sec:conclusions}

In this work, we revisited the gravitational and electromagnetic perturbations in a Reissner-Nordstr\"om black hole by means of the Newman Penrose formalism. In our analysis we include the sources that  cause the perturbations and 
discuss the particular case of
a charged perfect fluid falling radially into the black hole.
Using both, Maxwell equations for the Maxwell scalars and the Bianchi identities for the Weyl scalars we have 
found a system of coupled equations
for the gravitational and electromagnetic perturbations
without choosing a specific gauge.
A common practice to study electromagnetic and gravitational perturbations within the Newman Penrose formalism is to choose the so dubbed \emph{phantom gauge} (imposing  $\varphi_2^{(1)}=0$ ) since using this gauge one can obtain a sub-system of equations  for the perturbation of the Weyl scalars $\Psi_4^{(1)}$ and $\Psi_3^{(1)}$.
However, although convenient, this choice is not unique. 
In this work we have shown that a similar system of equations can be obtained for $\Psi_4^{(1)}$ and $\chi = 2\varphi_1 \Psi_3^{(1)}-3\Psi_2 \varphi_2^{(1)}$
which involves perturbations of the electromagnetic field $\varphi_2^{(1)}$ and perturbations of the electromagnetic part of the Weyl tensor $\Psi_3^{(1)}$. 
Our results thus opens up the possibility to explore electromagnetic and gravitational perturbations without any \emph{a priori} assumption on the value of any of the scalars.

It is remarkable that it is not possible to obtain a perturbation equation for $\Psi_3^{(1)}$ independently of $\varphi_2^{(1)}$ without choosing a particular gauge, only the combination given by $\chi$ can be determined via this formalism. 
However, this fact is not related with some physical properties of the fields since one can always obtain the fields by solving numerically the Einstein-Maxwell field equations and 
computing all the gravitational scalars at each time step. The actual physical meaning of such 
constraint in the Newman Penrose formalism is an ongoing work.

We also considered a dust-like charged fluid as a source of the perturbations. 
We used the spin weighted spherical harmonics as a basis to expand the functions and obtain a system of partial differential equations for the temporal and radial components, leaving all the angular dependence of the functions on the respective basis of spherical harmonics.
The resulting system of partial differential equations constitutes a hyperbolic system suitable to be solved numerically by standard means.
In this way, we see that we have a robust procedure which sets the basis for accurately determine the simultaneous  generation of gravitational and electromagnetic waveforms. A thorough study comparing amplitudes, frequencies, harmonic dependence and power between the gravitational and the electromagnetic/gravitational signals, for several fiducial values of the parameters will allow to determine correlations between these waveforms which not only will give a better understanding of the process, but in general might shed light in the multimessenger program as well, as long as it might be possible to extrapolate the correlations found in the system presented in this work, to other scenarios where signals of different interactions are generated.

Furthermore, our analysis can be used as a simple model to describe the correlation existing between electromagnetic and gravitational wave signals occurred during the accretion of charged matter around a compact object, since the frequency of the gravitational waves due to the quasinormal ringing of black holes of $(10$--$10^3)M_{\odot}$ with moderate charge lies within the range of sensibility of current ground base gravitational wave interferometers.

Finally, we remark that the studies made in a Reissner-Nordstr\"om spacetime
frequently give valuable insight in the Kerr geometry. The resulting relations arising from the interaction of the electromagnetic field of the matter with the charge of the black hole, might have a similitude with an interaction of the angular momentum of the accreting matter with the angular momentum of the black hole. The results and derivations presented in this work, might prove to be useful in the perturbation analysis generated by accreting rotating matter in a Kerr background. Such studies are currently underway.


\acknowledgments

This work was partially supported by 
DGAPA-UNAM through grants IN110218 and IN105920; by CONACyT Ciencia de Frontera Projects No. 376127 ``Sombras, lentes y ondas gravitatorias generadas por objetos
compactos astrof\'\i sicos", and No. 304001 ''Estudio de campos escalares con aplicaciones en cosmolog\'ia y astrof\'isica". Also by the European Union’s Horizon 2020 research and innovation (RISE) program H2020-MSCA-RISE-2017 Grant No. FunFiCO-777740. C. M. acknowledges support from PROSNI-UDG.  C. R.-L. acknowledges CONACYT scholarship. \\




\end{document}